\documentclass[a4paper,11pt]{article}
\pdfoutput=1 

\usepackage{jcappub} 
\usepackage[utf8]{inputenc}
\usepackage[T1]{fontenc}
\usepackage{graphicx,amssymb,amsmath,wasysym}
\usepackage{epsfig}
\usepackage{verbatim}
\usepackage{epstopdf}
\input{epsf.sty}
\usepackage{epsf}
\usepackage[dvipsnames]{xcolor}
\usepackage{color}
\usepackage{cancel}
\usepackage{babel}
\usepackage{array}
\usepackage{float}
\usepackage{booktabs}
\usepackage{multirow}
\usepackage{cancel}
\usepackage{soul}
\usepackage[normalem]{ulem}

\newcommand{\ba}{\begin{eqnarray}}
\newcommand{\ea}{\end{eqnarray}}
\newcommand{\bal}{\begin{align}}
\newcommand{\eal}{\end{align}}

\newcommand{\ds}{\phantom{s}} 
\newcommand{\iso}{\textrm{(iso)}}

\newcommand{\bfk}{\mathbf{k}}
\newcommand{\bfx}{\mathbf{x}}

\definecolor{blue4}{RGB}{0,0,143}

\newcommand{\be}{\begin{equation}}
\newcommand{\ee}{\end{equation}}

\begin{document}
\begin{flushright}
\end{flushright}

\title{ 
Structure formation in an anisotropic universe: Eulerian perturbation theory}

\author[a]{Juan P. Beltr\'an Almeida,}
\author[b]{Josu\'e Motoa-Manzano,}
\author[c]{Jorge Noreña,}
\author[d]{Thiago S. Pereira,}
\author[b]{C\'esar A. Valenzuela-Toledo.}

 \affiliation[a]{Universidad Nacional de Colombia, Facultad de Ciencias, Departamento de F\'isica,\\ Av. Cra 30 \# 45-03, Bogot\'a, Colombia}
 \affiliation[b]{Departamento de F\'isica, Universidad del Valle,
Ciudad Universitaria Mel\'endez, Santiago de Cali 760032, Colombia}
\affiliation[c]{Instituto de Física, Pontificia Universidad Católica de Valparaíso, Casilla 4950, Valparaíso,
Chile}
\affiliation[d]{Departamento de F\' isica, Universidade Estadual de Londrina,
Rod. Celso Garcia Cid, Km 380, 86057-970, Londrina, Paran\'a, Brazil.}

\emailAdd{jubeltrana@unal.edu.co, josue.motoa@correounivalle.edu.co, jorge.norena@pucv.cl, tspereira@uel.br,  cesar.valenzuela@correounivalle.edu.co.}

\vskip 0.5cm 

\abstract{We present an effective Eulerian description, in the non-relativistic regime, of the growth of cosmological perturbations around a homogeneous but anisotropic Bianchi I spacetime background.  We assume a small deviation from isotropy, sourced at late times for example by dark energy anisotropic stress. We thus derive an analytic solution for the linear dark matter density contrast, and use it in a formal perturbative approach which allows us to derive a second order (non-linear) solution. As an application of the procedure followed here we derive analytic expressions for the power spectrum and the bispectrum of the dark matter density contrast. The power spectrum receives a quadrupolar correction as expected, and the bispectrum receives several angle-dependent corrections. Quite generally, we find that the contribution of a late-time phase of anisotropic expansion to the growth of structure peaks at a finite redshift between CMB decoupling and today, tough the exact redshift value is model-dependent.}


\maketitle

\section{Introduction}

More than two decades have passed since the seminal discovery of the accelerated expansion of the Universe. Yet we still have no clue about the agent driving it, apart from the fact that it dominates the Universe’s energy budget and pushes galaxies away at an accelerated pace. 
The avalanche of observational data that followed its discovery, despite greatly improving our ability to narrow down the concordance model's parameter space, has also brought to the scene new observational anomalies \cite{Perivolaropoulos:2014lua,Schwarz:2015cma,Joudaki:2016kym,Hildebrandt:2016iqg,Planck:2019evm,DiValentino:2021izs,Perivolaropoulos:2021jda} whose origin and significance still eludes us. 
Thus, while we wait for future CMB and large scale structure surveys to shed light on the issue, it is crucial, on the theory side, that we test the robustness of the standard model assumptions that may have an impact on dark energy.

The standard model of cosmology is based on three main ingredients,\footnote{Four, if we consider the implicit assumption that the universe is topologically trivial -- see~\cite{Uzan:2016wji}.} all of which have direct impact on the description of dark energy. They have been explored to varying degrees in the literature in attempts to understand the physics behind it. 
First, we suppose that General Relativity is a valid description of gravitational interactions at cosmological scales, in which case dark energy is the result of a cosmological constant appearing in Einstein's field equations. 
If this hypothesis is false, one can in principle do away with dark energy by modifying gravitational interactions at large scales. This route has been vastly explored in the literature, and is often referred to as the \emph{Modified Gravity} approach for the accelerated expansion of the universe \cite{Nojiri:2006ri,Amendola:2006kh,Sotiriou:2008rp,Tsujikawa:2010zza,Joyce:2014kja,Planck:2015bue}. 
Second, we consider the standard model of particles and fields as a valid description of non-gravitational interactions in the universe. In this scenario, dark energy \emph{is} the vacuum energy of quantum fields, which has negative pressure and thus accelerates the expansion. However, the well-known discrepancy between the theoretical and observational values~\cite{Weinberg:2000yb} has forced theorists to search for alternative descriptions. The first obvious choice is to replace the vacuum for a dynamical entity, such as \emph{quintessence} \cite{Ratra:1987rm,Caldwell:1997ii} (see~\cite{Tsujikawa:2013fta,Yoo:2012ug} for a comprehensive discussion). 

The last ingredient in standard cosmology is the cosmological principle, i.e., the hypothesis that the universe is, at large scales, spatially homogeneous and isotropic. This line of research is usually hampered by the lack of exact cosmological models where spacetime symmetries can be ignored. Consequently, the physical interpretation of dark energy in this case depends on the proposal one chooses to follow. For example, soon after the discovery of the accelerated expansion it has been suggested that if we leave close to the center of a large void, we could explain the Hubble diagram without invoking dark energy~\cite{Celerier:1999hp,Tomita:2000jj,Caldwell:2007yu} (see also \cite{Marra:2011ct,Clarkson:2012bg} for reviews on this topic). It was later shown that void models capable of explaining the Hubble diagram appear to be severely constrained by the kSZ effect \cite{Zhang:2010fa}. However, this might not be the last word on the use of inhomogeneous models in cosmology \cite{Camarena:2021mjr}.

Still in connection with the cosmological principle, another possibility to test the physics of dark energy is to assume that it is described by an imperfect fluid \cite{Brevik:2005bj,Koivisto:2005mm,Mota:2007sz,Deffayet:2010qz}. In particular, the presence of a stress tensor in the fluid's description would result in a spatially anisotropic late-time universe \cite{Koivisto:2007bp,Koivisto:2008ig,Koivisto:2008xf,Thorsrud:2012mu,BeltranAlmeida:2019fou,Orjuela-Quintana:2020klr,Motoa-Manzano:2020mwe,Guarnizo:2020pkj,Orjuela-Quintana:2021zoe,Gomez:2021jbo}. This would not impair CMB constraints on isotropy, for two reasons. First, CMB strongly constrains the isotropy of the universe at \emph{early} times. Second, since the anisotropies induced by such stress appear nonlinearly in the expansion rate, but linearly in the fluid equations, it is possible that the anisotropies induced by dark energy remains subdominant in the Hubble diagram. On the other hand, they could be detectable in the growth of linear and nonlinear structures. Structure formation thus represents an important window to test models of anisotropic dark energy.

In the context of anisotropic dark energy models, Bianchi I metrics are the simplest and most interesting geometries to consider. Since they are cosmological solutions of Einstein field equations enjoying translation invariance, the eigenvectors of translation operators are given by the usual plane waves. Thus, the Fourier decomposition can be directly implemented in these models \cite{Pereira:2007yy}. In the limit of small stress, their main observational signature is the development of quadrupolar anisotropies affecting stochastic observables \cite{Marcori:2016oyn}. In particular, it was shown that a careful measurement of the $E$ and $B$ modes of the weak-lensing shear could be used to fully reconstruct the eigendirections of the expansion~\cite{Pitrou:2015iya,Pereira:2015jya,Adamek:2015mna}, and that future surveys such as Euclid and SKA are expected to constrain the spacetime shear to a percent level \cite{Pereira:2015jya}.

In a broader context, late-time tests of isotropy have focused mostly on type-Ia supernovae \cite{Kolatt:2000yg,Appleby:2009za,Koivisto:2010dr,Cai:2011xs,Appleby:2012as,Kalus:2012zu,Schucker:2014wca,Bengaly:2015dza}, although not without some criticism~\cite{Jimenez:2014jma}. Some tests have also used other late-time data~\cite{Bengaly:2015xkw,Andrade:2019kvl}. Here we contribute to this topic by considering the dynamics of linear and mildly non-linear dark matter perturbations in a Bianchi I geometry in the Newtonian limit. We assume a phenomenological description where the universe is isotropic prior to the dark matter regime (and thus described by the known FLRW solution), and slowly anisotropizes at late times due to the presence of a small shear tensor in the fluid describing dark energy. Assuming a small rate of anisotropic expansion, we use anisotropic perturbation theory \cite{Pitrou:2015iya} to derive the effective Eulerian equations for the dark matter density constrast and velocity potential in the Newtonian regime. In this regime, metric perturbations are small, spatial derivatives are large, and the dark matter density contrast is non-perturbative. We find a linear solution for the density contrast in this approximation, and use it in a formal perturbative scheme to derive the second order correction to the density contrast. Quite generally, we find that the contribution of a (hypothetical) dark stress tensor to the linear growth of perturbations will peak at a some redshift between the last scattering surface and today, though the precise redshift value depends on the specific model for dark energy. Similar approaches in the FLRW context, and second order calculations for the dynamic equations, taking into account relativistic corrections, can be found in the literature, see \emph{e.g.} \cite{Matarrese:1997ay,Bartolo:2005kv,Pitrou:2008hy,Castiblanco:2018qsd} and references therein.

We start by recalling the basic features of Bianchi I spacetimes in Section~\ref{sec:PT}, where we also explain the details of the Newtonian approximation in this setup. This section ends with the derivation of our first main result, given by Eqs.~(\ref{eq:delta}) and (\ref{eq:theta}). In Section~\ref{solutions} we derive the exact (linear) and approximate (non-linear) solutions for the dark matter density contrast in the small stress approximation, and give the corresponding Fourier kernels. We present the two- ant three-point correlation functions in Section~\ref{sec:correlations}, where we also comment on the consistency relation in this case. Finally, we conclude and discuss some perspectives in Section~\ref{sec:conclusions}.

\section{Perturbation theory}\label{sec:PT}
In this section we define the metric and fluid variables necessary for the description of the gravitational clustering process in an anisotropic background. We also define the small shear and weak-field approximations that we shall adopt. After that, we write the expressions, in the Newtonian limit, for fluid and Einstein's equations. Similar approaches and second order calculations for the dynamic equations, taking into account relativistic corrections can be found in the literature, see \emph{e.g.} \cite{Matarrese:1997ay, Bartolo:2005kv,Castiblanco:2018qsd} and references therein.

\subsection{Background geometry and dynamics}\label{subsec:background}
The physical and mathematical aspects of Bianchi-I universes have been vastly explored in the literature, and good introductions to the subject can be found in standard textbooks~\cite{PeterUzan,plebanski2006,ellis2012}. The starting point is the Bianchi-I geometry, described by the line element
\be\label{BI-metric}
ds^2 = a^2(\eta)[-d\eta^2+\gamma_{ij}(\eta)dx^i dx^j]\,.
\ee
It differs from the standard Friedmann-Lemaître-Robertson-Walker (FLRW) metric insofar as, in the former, the metric of constant-time hypersurfaces is time-dependent. This leads to an anisotropic expansion of the universe, which is fully characterized by the shear tensor
\be\label{def-shear}
\sigma_{ij} = \frac{1}{2}\gamma'_{ij}\,,
\ee
where a prime denotes derivative with respect to conformal time. Spatial indices are manipulated with $\gamma_{ij}$ and its inverse, and since both are time-dependent, we have that
\be\label{shear-ids}
2\sigma^{ij}=\gamma^{ik}\gamma^{jl}\gamma_{kl}'\neq(\gamma^{ij})' = -2\sigma^{ij}\,. 
\ee

Constant-time hypersurfaces of Bianchi-I universes are spatially flat and have trivial isometries\footnote{That is, spatial translations are given by straight lines.}. This means that comoving cartesian coordinates $\bfx=\{x^j\}$ are constant, so that any function $f(\eta,\bfx)$ can be Fourier-decomposed as usual:
\begin{align}
f(\eta,\bfx) & = \int \frac{d^3\bfk}{(2\pi)^3}\, f(\eta,\bfk)e^{i\bfk\cdot\bfx}\,,\\
f(\eta,\bfk) & = \int d^3\bfx\,f(\eta,\bfx)
e^{-i\bfk\cdot\bfx}\,.
\end{align}
Here, $\bfk\cdot\bfx=k_jx^j$ is constant, which means that the comoving co-vector $\bfk=\{k_j\}$ is also constant. However, $k^i=\gamma^{ij}k_j$ is time-dependent, so that dot products like $\bfk_1\cdot\bfk_2$ have an implicit time-dependence through the spatial metric. \emph{In what follows, boldface letters will always represent constant (comoving) quantities}. Before moving on, it is appropriate to define the contraction of the shear with two unit Fourier vectors:
\be\label{sigmapar}
\sigma_\parallel(\eta,\bfk) \equiv \sigma_{ij}(\eta) \hat{k}^i\hat{k}^j\,,
\ee
which will later appear in the dynamical equations.

In anisotropic cosmologies, it is important to distinguish the overal expansion of the universe from the (volume-preserving) anisotropies of the geometry. Thus, the scale factor $a$ in the metric~\eqref{BI-metric} describes the overall expansion, whereas the metric $\gamma_{ij}$, despite being time-dependent, has unit determinant. This last condition is easily implemented by using Misner's parameterization of the metric~\cite{Misner:1969hg}: 
\be\label{misnermetric}
\gamma_{ij} = [e^{2\lambda}]_{ij}\,,\qquad \textrm{det}(\gamma) = 1\,.
\ee
In the eigenbasis of the expansion, where $\gamma_{ij}=\textrm{diag}(e^{2\lambda_1},e^{2\lambda_2},e^{2\lambda_3})$, this implies that $\sum_i\lambda_i=0$. In particular, since $\sigma_{ij}=\gamma'_{ij}/2$, we find 
that the shear is traceless:
\be
\sigma^i_{\ds i} = \sum_i\lambda'_i=0\,.
\ee

Moving to the dynamics, the cosmological scenario we have in mind is that of an early (isotropic) dark matter dominated universe which, at late times, slowly enters into an anisotropic dark energy dominated phase. The appropriate energy-momentum tensor is thus of the form $T^{\mu\nu}=T_{m}^{\mu\nu}+T_{de}^{\mu\nu}$, where:
\begin{align}
T^{\mu\nu}_m & = \rho_m u^\mu u^\nu\,, \label{tmunudm}\\
T^{\mu\nu}_{de} & = (\rho_{de}+p_{de})u^\mu u^\nu + p_{de} g^{\mu\nu} + \pi_{de}^{\mu\nu}\label{tmunude}\,,
\end{align}
and where $\pi_{de}^{\mu\nu}$ is a homogeneous stress tensor satisfying $u_\mu\pi_{de}^{\mu\nu}=0=g_{\mu\nu}\pi_{de}^{\mu\nu}$. 
For simplicity, we shall drop the subscript and write it simply as $\pi^{\mu\nu}$, since it is the only stress tensor we shall be considering in this work. The dynamics of the expansion is then given by Einstein and continuity equations, which in this particular case become\footnote{We adopt $8\pi G=1$.}
\begin{align}
3{\cal H}^2 & = (\rho_m+\rho_{de}) a^2 + \frac{1}{2}\sigma^2\,,\label{Friedmann_eq00}\\
(\sigma^i_{\ds j})'+2{\cal H}\sigma^i_{\ds j} & = a^2\pi^i_{\ds j}\,, \\
\rho'_m+3{\cal H}\rho_m & = 0\,,\\
\rho'_{de}+3{\cal H}(\rho_{de}+p_{de}) & = -\sigma_{ij}\pi^{ij}\label{rhode_eq}\,.
\end{align}
where $\sigma^2\equiv\sigma^{ij}\sigma_{ij}$. As usual, the system of equations is not closed until we have equations of state for $p_{de}$ and, in our case, also for $\pi_{ij}$. While we shall follow the standard procedure of adopting a constant equation for the former, a great deal of information can be obtained without making any assumption about the latter. Indeed, by integrating the equation for the shear,
\be\label{eq:shear}
\sigma_{\ds j}^{i} =\left(\frac{a_i}{a}\right)^2{}^{(0)}\sigma_{\ds j}^{i}+\frac{1}{a^{2}}\int_{a_i}^{a}s^{3}\frac{\pi_{\ds j}^{i}}{\cal H}\,ds\,,
\ee
we see that it is composed of early and late-time contributions.  The early solution is highly constrained at high redshifts by CMB data: $^{(0)}\sigma/{\cal H}\lesssim 10^{-10}$ \cite{martinez1995delta,Maartens:1995hh,Saadeh:2016sak}. Since this solution corresponds to a decaying mode, it is completely negligible at small redshifts. Observational constraints on the late-time solution, on the other hand, while much weaker in comparison to those of CMB, still point to a nearly isotropic dark-energy component~\cite{Bengaly:2015xkw,Andrade:2019kvl}. This justifies the adoption of a \emph{small shear} approximation,
\be\label{small-shear}
\frac{\pi}{\cal H}\sim\frac{\sigma}{\cal H} \ll 1\,,
\ee
while looking for approximate solutions for the growth of structure. Throughout this paper we will work up to linear order in $\sigma/{\cal H}$. Besides being physically motivated, this approximation has the benefit of making the background dynamics the same as the isotropic one, since Eqs.~(\ref{Friedmann_eq00}) and (\ref{rhode_eq}) are only affected at second order in $\sigma/\mathcal{H}$. Indeed, in this case the shear dynamics is essentially that of a infinite wavelength gravitational wave~\cite{Pontzen:2010eg,Pereira:2019mpp}. Having this in mind, in the remainder of this work we shall adopt
\be
p_{de} = p_\Lambda = -\rho_\Lambda = -\rho_{de}\,,
\ee
which solves Eq.~(\ref{rhode_eq}) up to first order in $\sigma/\mathcal{H}$. In section~\ref{solutions} we shall look for approximate solutions, departing from the FLRW ones, but treating $\sigma_{ij}$ and $\pi_{ij}$ as small sources.

\subsection{Newtonian limit and perturbed variables}
In this work we want to study the growth of structures in an anisotropically expanding universe. Since cold dark matter particles are non-relativistic, and non-linearities develop on scales below the Hubble radius, it is enough as a first approach to consider the Newtonian limit of the relativistic field equations which, for a Bianchi-I setup, where derived in Ref.~\cite{Pitrou:2015iya}. We thus adopt the so-called weak-field approximation~\cite{Green:2010qy,Green:2011wc,Brustein:2011dy,Kopp:2013tqa}, which allows us to probe structures even at very non-linear scales, provided that velocities are still small~\cite{Castiblanco:2018qsd}. In a nutshell, it consists in treating metric perturbations as small, but space derivatives as large. In order to justify this approximation we can examine the Newtonian equations for the gravitational potential and velocity perturbations, which are given by
\[
k^2\phi=-\frac{3}{2}a^2H^2\delta\,,\qquad\text{and}\qquad v_i = -\frac{k_i\phi}{aH}\,. 
\]
From CMB observations we know that $\phi\sim {\cal O}(10^{-5})$ at horizon scales where $k\sim aH$. It follows that, at these scales, velocities are small: $v_i\sim {\cal O}(10^{-5})$. On the other hand, at small scales where $aH/k\sim{\cal O}(10^{-3})$ and $\delta\sim{\cal O}(1)$, velocities are also known to be small, $v_i\sim{\cal O}(10^{-3})$, which implies again that $\phi\sim{\cal O}(10^{-5})$. We thus conclude that $\phi\sim{\cal O}(10^{-5})$ at all scales from the horizon down to the small scales where non-linearities are important and velocities are non-relativistic. Since spatial derivatives become important at small scales, we adopt a perturbative expansion scheme in the parameter
\be
\epsilon \equiv \left(\frac{aH}{k}\right)^2\,.
\ee
Note that this parameter, which controls the size of \emph{stochastic} variables, is independent of the approximation (\ref{small-shear}) on the \emph{geometric} shear, so that we are conducting a two-parameter perturbative scheme: $\epsilon$ for the size of stochastic variables and $\sigma/{\cal H}$ for the deviation of the background metric from isotropy. Keeping track of powers in $\sigma/{\cal H}$ is straightforward. 
We thus make the following ansatz: stochastic metric fluctuations are at most of order $\epsilon$, perturbations of the fluid's velocity are at most of order $\epsilon^{1/2}$, and spatial derivatives are of order $\epsilon^{-1/2}$. The fact that this ansatz leads to a consistent series in powers of $\epsilon$ can be checked a posteriori using the perturbed Einstein and fluid equations. That being said, \emph{we define the Newtonian limit as the leading order equations in the parameter} $\epsilon$. Table~\ref{table-weak-field} summarizes the order of metric and fluid perturbations that we shall later encounter. 
\begin{table}[H]
\begin{centering}
\begin{tabular}{ccc}
\toprule 
 & Variable & $n$ in ${\cal O}(\epsilon^{n})$\tabularnewline
\midrule
\midrule 
Derivatives & $\partial_{i}/aH$ & $-1/2$\tabularnewline
\midrule 
\multirow{4}{*}{Fluid} & $\rho,\,\delta,\,\theta$ & $0$\tabularnewline
\cmidrule{2-3} \cmidrule{3-3} 
 & $v$ & $1$\tabularnewline
\cmidrule{2-3} \cmidrule{3-3} 
 & $v_{i}$ & $1/2$\tabularnewline
\midrule 
\multirow{4}{*}{Metric}  
 & $\phi,\,\psi$ & $1$\tabularnewline
\cmidrule{2-3} \cmidrule{3-3} 
 & $\omega_{i}$ & $3/2$\tabularnewline
\cmidrule{2-3} \cmidrule{3-3} 
 & $\tau_{ij}$ & $1$\tabularnewline
\bottomrule
\end{tabular}
\par\end{centering}
\caption{Order of fluid and metric perturbations in the weak-field approximation.}\label{table-weak-field}
\end{table}

The definition of perturbations in a Bianchi-I setup is straightforward, and for the details we point the reader to Ref.~\cite{Pereira:2007yy}. We shall be working in Newtonian gauge, where the gauge freedom is completely fixed. At linear order, the perturbed line element in conformal time is given by
\be\label{backgd_metric} 
ds^{2}=a^{2}\left[-(1+2\phi)d\eta^{2}+2\omega_{i}dx^{i}d\eta+\left(\gamma_{ij}+h_{ij}\right)dx^{i}dx^{j}\right]\,,
\ee
where
\[
h_{ij}\equiv-2\psi\left(\gamma_{ij}+\frac{\sigma_{ij}}{{\cal H}}\right)+2\tau_{ij}\,,
\]
and
\[
\partial^{i}\omega_{i}=\tau^{i}_{\ds i}=0=\partial^{i}\tau_{ij}\,.
\]
The components of the inverse metric at order $\mathcal{O}(\epsilon)$ are
\be
g^{00}=-a^{-2}\left(1-2\phi\right)\,,\quad g^{0i}=a^{-2}\omega^{i}\,,\quad g^{ij}=a^{-2}(\gamma^{ij}-h^{ij})\,.
\ee

In order to proceed we need a prescription for the matter content. As explained above, the scenario we have in mind is the closest possible to a $\Lambda$CDM universe, except for the fact that we allow dark energy --- or more precisely, the cosmological constant --- to develop late-time anisotropy through a (homogeneous) stress tensor $\pi_{ij}$. Since $\pi_{ij}$ is presumably small, in the sense of approximation~\eqref{small-shear}, we ignore its perturbations, and thus the perturbations of dark energy altogether. We thus focus on the perturbations of the dark matter component, which is taken to be an irrotational presureless perfect fluid with a dust-like energy-momentum tensor~\eqref{tmunudm}. Thus we expand the dark matter density in terms of a density contrast $\delta$ as usual
\be\label{def-delta}
\rho_m \rightarrow \rho_m+ \delta\rho_m\equiv \rho_m(1+\delta) \,.
\ee
Since $\delta$ is expected to receive a contribution from the Laplacian of the metric perturbations, it is not suppressed by $\epsilon$, i.e. $\delta \sim \mathcal{O}(\epsilon^0)$.\footnote{In FLRW, the resulting equations to leading order in the weak field approximation are fully non-linear equations in $\delta$ that correspond to the usual Newtonian standard perturbation theory equations.}

The fluid four-velocity can be writen as ${u^\mu = a^{-1}(1+\delta u^0,v^i)}$, where $v^i$ is given by the sum of longitudinal and transverse parts, as usual. However, as we show in Appendix~\ref{transversev}, to order $\mathcal{O}(\epsilon)$ and in the small shear limit, the transverse part of the dark matter velocity can be neglected. Thus, the spatial velocity can be completely characterized by one scalar function, say, the fluid divergence $\theta=\partial_i v^i$, or, formally:
\be\label{vi-theta}
v^i=\frac{\partial^i \theta}{\nabla^2}\,.
\ee
Together with the dark matter contrast $\delta$, the divergence fully specify the fluid's perturbations. It remains to normalize the four-velocity $u^\mu$. To order $\mathcal{O}(\epsilon)$, we find
\be
u^{\mu}=a^{-1}\left(1-\phi+\frac{1}{2}v^{2},v^{i}\right),
\ee
where $v^2 = \gamma_{ij}v^i v^j$. 

Finally, let us check that the ansatz given in Table~\ref{table-weak-field} is consistent with the relativistic equations to leading order in the weak-fied approximation. Using the $0i$ perturbed Einstein equation given by Eq.~C11 of Ref.~\cite{Pereira:2007yy}, we have that
\begin{align}\label{0ieq}
\sigma^j_{\ds i}\partial_j\left[\phi+\psi+\left(\frac{\psi}{\cal H}\right)'\right]\nonumber &+ \frac{1}{2}\nabla^2\omega_i-2\sigma^{jk}\partial_j\tau_{ik}\\ 
& + \partial_i\left[-\frac{\sigma^2}{\cal H}\psi-2(\psi'+{\cal H}\phi)+\sigma^{jk}\tau_{jk}\right] = a^2\rho_m u^0u_i\,.
\end{align}
The scalar part of this expression can be extracted with the operator $\partial^i$. The resulting equation is
\[
\sigma^{ij}\partial_i\partial_j\left[\phi+\psi+\left(\frac{\psi}{\cal H}\right)'\right]+\nabla^2\left[-\frac{\sigma^2}{\cal H}\psi-2(\psi'+{\cal H}\phi)+\sigma^{jk}\tau_{jk}\right] = a\partial^i[\rho_m u^0v_i]\,.
\]
where we have used $u^i=a^{-1}v^i$. Since the lhs is of order $\mathcal{O}(\epsilon^0)$, we find that $v_i\sim\mathcal{O}(\epsilon^{1/2})$, in agreement with our ansatz. In particular, this implies that, to order $\epsilon$, the index in $v_i$ can be manipulated with the background metric. 

The vector part of Eq.~\eqref{0ieq} can be extracted with the operator $P^i_{\ds j} = \delta^i_j-\partial^i\partial_j/\nabla^2$. It gives
\[
P^i_{\ds l}[\nabla^2\omega_i] - 4P^i_{\ds l}[\sigma^{jk}\partial_j\tau_{ik}] = -2P^i_{\ds l}\left[\sigma^j_{\ds i}\partial_j\left(\phi+\psi+\left(\frac{\psi}{\cal H}\right)'\right)\right]\,,
\]
where we have used the fact that $v^i$ has no transverse part (see Appendix~\ref{transversev}). The rhs of this equation is of order $\mathcal{O}(\epsilon^{1/2})$, and since ${\nabla^2\sim\mathcal{O}(\epsilon^{-1}})$, we find that $\omega_i\sim\mathcal{O}(\epsilon^{3/2})$ and $\tau_{ij}\sim\mathcal{O}(\epsilon)$, in agreement with Table~\ref{table-weak-field}.

\subsection{Perturbed equations}

We now proceed by explicitly writing the dark-matter fluid equations in the Newtonian limit. Conservation of stress-energy for matter leads to the relevant equations, which are the continuity and Euler equations, given respectively by
\begin{align}
\nabla_\mu[\rho_m(1+\delta)u^\mu] & = 0\,, \\
u^\nu\nabla_\nu u^\mu & = 0\,.
\end{align}
Since these equations contain at most one spatial derivative, we will need the Christoffel symbols to order $\mathcal{O}(\epsilon^{1/2})$; these are given in Appendix~\ref{christoffels}. Starting with the continuity equation, a straightforward computation leads to
\begin{equation}
\delta'+\partial_i\left[(1+\delta)v^i\right]=0\,.\label{eq:continuity_nr}
\end{equation}
We conclude that, to order $\mathcal{O}(\epsilon^0)$, the shear will not change the continuity equation directly. The Euler equation, on the other hand, is of order ${\cal O}(\epsilon^{1/2})$, and at this order it is modified by the shear:
\be
\label{eq:full_euler}
(v^i)'+{\cal H}v^i + v^j\partial_j v^i + \partial^i\phi + 2\sigma^i_{\ds j}v^j = 0\,,
\ee
or, in terms of the divergence $\theta$,\footnote{In terms of $v_i$, Euler equation reads $v_i'+{\cal H}v_i + v^j\partial_j v_i + \partial_i\phi = 0$. If we now take the divergence $\partial^i=\gamma^{ij}\partial_j$ on both sides, and recall the identity~\eqref{shear-ids}, one again recovers Eq.~\eqref{theta_sigma}.}
\be\label{theta_sigma}
\theta'+{\cal H}\theta + \partial_i(v^j\partial_j v^i) + \nabla^2\phi + 2\sigma_{ij}\partial^i v^j = 0\,.
\ee

Equations (\ref{eq:continuity_nr}) and \eqref{theta_sigma} are coupled to each other and also to the metric perturbations. However, the dependence on the latter can be eliminated with the help of Einstein's field equations. We wish to write expressions which are arbitrarily non-linear in $\delta$, but valid in the non-relativistic limit up to order $\mathcal{O}(\epsilon^0)$. At this order, and without further perturbative expansions, the Einstein tensor contains a finite number of terms: Each metric perturbation is at most $\mathcal{O}(\epsilon)$, and the Einstein tensor contains two derivatives of the metric. So there will be terms of order $\mathcal{O}(\epsilon^0)$ built from a single metric fluctuation and two spatial derivatives. For this reason, we can recycle the \emph{linearized} expressions for the Einstein tensor, and write Einstein equations which will be valid at order $\mathcal{O}(\epsilon^0)$ and are non-linear in $\delta$. Thus, for example, from the 00 perturbed equation in~\cite{Pereira:2007yy} we find the following
\be
-3{\cal H}^{2}+\frac{\sigma^{2}}{2}-2\nabla^{2}\psi+\frac{\sigma^{ij}}{{\cal H}}\partial_{i}\partial_{j}\psi  = -a^{2}[\rho_{m}(1+\delta)+\rho_{\Lambda}]\,,
\ee
where we have used $\delta \rho_{\Lambda} = 0 $ and the definition~\eqref{def-delta}.  Using the (background) equation $-3{\cal H}^{2}+\sigma^{2}/2=-a^{2}\left(\rho_{m}+\rho_{\Lambda}\right)$, this simplifies to
\be\label{eq:psi}
2\nabla^{2}\psi-\frac{\sigma^{ij}}{{\cal H}}\partial_{i}\partial_{j}\psi=a^{2}\rho_{m}\delta\,.
\ee
The potential $\psi$, on the other hand, is related to $\phi$ through the trace of the $ij$ Einstein equation. To order $\mathcal{O}(\epsilon^0)$ we find~\cite{Pereira:2007yy}
\be
2\nabla^2(\phi-\psi)-\frac{\sigma^{ij}}{\cal H}\partial_i\partial_j\psi - \left(2{\cal H}'+{\cal H}^2+\frac{\sigma^2}{2}\right) = -a^2\rho_\Lambda\,.
\ee
By manipulating the background equations one can show that the term in parenthesis gives $a^2\rho_\Lambda$, so that
\begin{equation}
2\nabla^{2}(\psi-\phi)-\frac{\sigma^{ij}}{{\cal H}}\partial_{i}\partial_{j}\psi=0\,.\label{eq:psi-phi}
\end{equation}
Combining Eqs.~\eqref{eq:psi} and \eqref{eq:psi-phi}, we then arrive at
\be\label{eq:Poisson}
\nabla^2\phi = \frac{3}{2}\Omega_m{\cal H}^2\delta\,,
\ee
where we have introduced $\Omega_m\equiv a^2\rho_m/3{\cal H}^2$. While this is formally the same as the Poisson equation in Newtonian gravity, one should recall that the Laplacian $\nabla^2=\gamma^{ij}\partial_i\partial_j$ is now time-dependent. 
Moreover, the potential $\psi$ and $\phi$ are non longer equal in the presence of the shear. Indeed, from eq. (\ref{eq:psi-phi}) in Fourier space, we find that
\be
\psi=\phi\left(1+\frac{\sigma_{\parallel}}{2{\cal H}}\right)+{\mathcal O}\left(\left(\sigma/\cal H\right)^2\right)\,.
\ee
Finally, using~\eqref{eq:Poisson}, the fluid equations can be rewritten as a consistent system of coupled differential equations:
\begin{align}
\delta'+ \theta(1 +\delta) + v^i\partial_i\delta & = 0\,,\label{eq:delta} \\
\theta' + {\cal H}\theta + \partial_i(v^j\partial_j v^i)+\frac{3}{2}\Omega_m{\cal H}^2\delta& = -2\sigma^{ij}\partial_i v_j\label{eq:theta}\,,
\end{align}
which is closed by relation~(\ref{vi-theta}). This set of equations is one of the main results of this work. It describes the growth of structure in an anisotropic Bianchi I universe, at all orders in perturbation theory (as long as the fluid approximation holds). They are simply the non-relativistic limit of the Einstein and stress-energy conservation equations, and are analogous to the standard perturbation theory equations for the isotropic universe.

Let us briefly discuss the application of the Effective Field Theory (EFT) of Large Scale Structure (LSS) \cite{Baumann:2010tm, Carrasco:2012cv} to this case. It will induce counterterms appearing as additional terms in equations \eqref{eq:delta}-\eqref{eq:theta}. Since these are expected to be small corrections to the equations, and since we take $\sigma^{ij}$ to be small, we expect new anisotropic counterterms to be negligible. Thus, it would be a reasonable approximation to include the usual \emph{isotropic} counterterms in equations \eqref{eq:delta}-\eqref{eq:theta}. Here, we only perform tree level calculations, so we don't need to include those terms.

\section{Solutions}\label{solutions}
Let us now investigate the solutions of the system (\ref{eq:delta})-(\ref{eq:theta}) in a universe containing dark matter and anistropic dark energy. Because relation~(\ref{vi-theta}) is nonlocal, it is appropriate to work in Fourier space, where it reads
\be
v^i(\eta,\bfk) = -i\frac{k^i}{k^2}\theta(\eta,\bfk) \,,
\ee
and from which we can infer that $\theta(\eta,\bfk) = i k_i v^i(\eta,\bfk)$. With this in mind, and recalling~(\ref{sigmapar}), the Fourier space representation of the fluid equations reads
\ba\label{eomkdelta} \nonumber
&&\delta' (\eta,{\bf{k}})  + \theta (\eta,{\bf{k}})  \\
&& \qquad   \qquad \qquad \qquad=- \int [{\rm d}k]^2  (2\pi)^3\delta_D ({\bf k}-{\bf k}_{12})\alpha (\eta,{\bf k}_1, {\bf k}_2)\theta (\eta,{\bf{k}}_1)\delta (\eta, {\bf{k}}_2)\,, \\ \nonumber
&& \theta' (\eta, {\bf{k}}) +  \left[{\cal{H}} +2\sigma_{\parallel}(\eta,\bfk)\right]\theta (\eta,{\bf{k}}) + \frac{3}{2}{\cal{H}}^2\Omega_m \delta (\eta, {\bf{k}}) \\ \label{eomktheta}
&& \qquad   \qquad \qquad \qquad = - \int   [{\rm d}k]^2 (2\pi)^3\delta_D ({\bf k}-{\bf k}_{12})\beta (\eta,{\bf k}_1, {\bf k}_2)\theta (\eta,{\bf{k}}_1)\theta (\eta,{\bf{k}}_2)\,,
\ea
where $\delta_D$ is Dirac's delta function and where we have introduced the notation
\begin{align}
[{\rm d}k]^n & \equiv \frac{d^3\bfk_1}{(2\pi)^3}\cdots \frac{d^3\bfk_n}{(2\pi)^3}\,, \\
\bfk_{ab} & \equiv\bfk_a+\bfk_b\,.
\end{align}
The nonlinearity of the system is expressed by the mode coupling functions $\alpha$ and $\beta$, which have the same formal definition as their FLRW counterparts (see, \emph{e.g.},~\cite{Bernardeau:2001qr})
\be\label{alpha-beta}
\alpha (\eta,{\bf k}_1, {\bf k}_2)\equiv \frac{{\bf k}_{12}\cdot {\bf k}_1}{k_1^2}\qquad \text{and}\qquad\beta  (\eta,{\bf k}_1, {\bf k}_2) \equiv \frac{({\bf k}_{12})^2 {\bf k}_1\cdot {\bf k}_2 }{2 k_1^2 k_2^2},
\ee 
but which now are time-dependent since they involve $\gamma^{ij}$ in the dot products.

Equations \eqref{eomkdelta}-\eqref{eomktheta} constitute, to leading order in the weak-field and small shear approximations, the starting point for a perturbative approach for the dynamics of the gravitational clustering in an anisotropic background. From these equations we can see that a late-time phase of anisotropy induces quadrupolar corrections in  the fluid divergence and density contrast. This happens to be a general signature of models with anisotropic expansions since, in a spherical basis, the quantity~(\ref{sigmapar}) is essentially a quadrupole.\footnote{In a spherical basis, $\sigma_\parallel(\eta,{\bf k}) = \sum_m c_m(\eta,k)Y_{2m}(\hat{\bf{k}})$. Even multipoles of higher order would result from higher corrections to the small shear expansion, whereas odd multipoles are forbidden by parity symmetry of Bianchi-I spacetimes~\cite{Pereira:2015pga}.} Since we are treating $\sigma_\parallel$ as small, this structure propagates linearly to the two and three point correlators of the density contrast, as we shall see.

\subsection{Linear solutions}
\label{sect:linear}
By definition, the coupling terms are neglected in the linear approximation, and equation~(\ref{eomkdelta}) gives $\delta'_1=-\theta_1$, where from now on $\delta_1$ denotes the linear density contrast, and $\theta_1$ the linear velocity divergence. Plugging this result back in equation~(\ref{eomktheta}), we find 
\be\label{eomkdelta_linear}
\delta''_1(\eta,{\bf{k}}) + {\cal{H}}\delta'_1(\eta,{\bf{k}})-\frac{3}{2}{\cal{H}}^2\Omega_m \delta_1(\eta,{\bf{k}}) = - 2\sigma_\parallel(\eta,\bfk)\delta'_1(\eta,{\bf{k}})\,. 
\ee
In order to solve this equation we need to find the time dependence of $\sigma_\parallel$, which in turn requires us to solve the background equation for $\sigma_{ij}$ in some basis (\emph{e.g.}, the eigenbasis of the expansion). However, an inspection of (\ref{sigmapar}) shows that $\sigma_\parallel(\eta,\bfk)$ is in general not a separable function of its arguments, and this approach would lead us nowhere.\footnote{Note that even if we choose $\hat{k}^i=\delta^i_3$ initially -- in which case $\sigma_\parallel=\sigma_{33}(\eta)$ -- the anisotropy of the expansion would not preserve this choice.} On the other hand, since the shear is presumably small, we can treat the rhs of (\ref{eomkdelta_linear}) as an external source, and use standard techniques to solve nonhomogeneous differential equations. From the isotropic theory (see, \emph{e.g.},~\cite{PeterUzan}), we know that (\ref{eomkdelta_linear}) admits two linearly independent \emph{homogeneous} solutions of the form
\be
D_{+}(\eta)A(\bfk)\,,\qquad\text{and}\qquad D_{-}(\eta)B(\bfk)\,,
\ee
where the time-dependent functions depend on the matter content of the universe. Thus, the general solution to (\ref{eomkdelta_linear}) will be formally given by
\begin{align}\label{deltaaniso-eds}
\delta_1(\eta,\bfk) = D_{+}(\eta)A(\bfk) & + D_{-}(\eta)B(\bfk)\nonumber \\
& -2\int^\eta\frac{D_{-}(\eta)D_{+}(y)-D_{+}(\eta)D_{-}(y)}{W[D_{+},D_{-}](y)}\sigma_\parallel(y,\bfk)\delta_1^{\iso'}(y,\bfk)dy\,,
\end{align}
where $W[D_+,D_-]=D_{+}D'_{-}-D'_{+}D_{-}$ is the Wroskian of the (homogeneous) solutions and $\delta_1^{\iso}$ is the linear density contrast in isotropic universes:
\be\label{deltaiso}
\delta_1^{\iso}(\eta,\bfk) = D_{+}(\eta)A(\bfk) + D_{-}(\eta)B(\bfk)\,.
\ee
Since the integral in (\ref{deltaaniso-eds}) is indeterminate, this solution is defined up to an arbitrary constant. We can nonetheless fix this constant uniquely by demanding that $\delta_1$ equals $\delta_1^\iso$ at some initial time $\eta_i$, which is consistent with our hypothesis that the early universe is isotropic.

As an example, let us see how this solution works in the simple case of a matter dominated universe, where $\Omega_m=1$, $a=(\eta/\eta_0)^2$ and ${\cal H}=2/\eta$. As is well-known, in this case $D_{+}\propto\eta^2$ and $D_{-}\propto\eta^{-3}$, so that $W\propto-5\eta^{-2}$. A simple algebra leads us to
\be
\delta_1(\eta,\bfk) = A(\bfk)\eta^2\left[1 - \frac{4}{5}\int_{\eta_i}^\eta\left(\frac{\eta^5-y^5}{\eta^5}\right)\sigma_\parallel(y,\bfk) dy\right],
\ee
where we have neglected the decaying mode. Thus, the linear density contrast splits into the usual Newtonian term, which grows linearly with the scale factor, plus a correction linear in the shear. Note also that the initial condition $\delta_1(\eta_i) = \delta_1^{\iso}(\eta_i)$ has been explicitly adopted.

The solution for $\delta_1$ in the interesting case of a universe containing matter and dark-energy requires a bit more work. For this case, it is easier to adopt the scale factor as a time parameter in~(\ref{eomkdelta_linear}). Representing derivatives with respect to $a$ by a dot, a straightforward computation gives
\be \label{delta2ndDegreelcdm}
\ddot{\delta}_1(a,\bfk)+\left(\frac{\dot{H}}{H}+\frac{3}{a}\right)\dot{\delta}_1(a,\bfk)-\frac{3}{2}\frac{\Omega_m^0}{a^5}\left(\frac{H_0}{H}\right)^{2}\delta_1(a,\bfk)
=-\frac{2\sigma_{\parallel}(a,\bfk)\dot{\delta}_1(a,\bfk)}{a^2H}\,.
\ee
The homogeneous solution of this equation is again given by a linear combination of time-dependent functions $D_{+}(a)$ and $D_{-}(a)$ defined as
\be
\begin{split}\label{Dplusminus}
D_{+}(a) & = \frac{5}{2}\Omega_m^0\frac{H(a)}{H_0}\int^a_{0}\frac{ds}{[sH(s)/H_0]^3}\,, \\
D_{-}(a) & = \frac{H(a)}{H_0}\,.
\end{split}
\ee
We recall that the background quantities in the above expressions are given by standard FLRW equations, since the evolution is isotropic at linear order in the shear. To linear order in $\sigma_\parallel$, the general solution is
\begin{align}\label{deltaaniso-lcdm}
\delta_1(a,\bfk) = D_{+}(a)A(\bfk) & + D_{-}(a)B(\bfk)\nonumber \\
& -2 \int^a\frac{D_{-}(a)D_{+}(s)-D_{+}(a)D_{-}(s)}{W[D_{+},D_{-}](s)}\frac{\sigma_\parallel(s,\bfk)\dot{\delta}_1^{\iso}(s,\bfk)}{s^2H(s)} ds\,,
\end{align}
 where the Wronskian now reads
\be
W[D_{+},D_{-}] = -\frac{5\Omega_m^0}{2s^3}\frac{H_0}{H(s)}\,,
\ee
and $\delta_1^\iso(a,\bfk)=D_+(a)A(\bfk)+D_-(a)B(\bfk)$. But given that the decaying mode in $\delta_1^\iso$ will eventually be discarded (otherwise it spoils early-time physics), let us drop it for once and write
\be\label{delta1iso}
\delta_1^\iso(a,\bfk)\equiv D_+(a)\delta_l(\bfk)\,,
\ee
which is a more standard notation. Finally, plugging this expression in (\ref{deltaaniso-lcdm}) and rearranging some terms, the linear solution in a $\Lambda$CDM universe can be written as\footnote{Note that, since we are working to first order in $\sigma_{ij}$, we are allowed to remove $\hat{k}^i$ from the time integrals.}
\begin{align}\label{delta-linear-lcdm}
\delta_1(a,\bfk)  =    D_{+}(a)\left(1 + Q_{ij}(a)\hat{k}^i \hat{k}^j\right)\delta_l(\bfk)\,,
\end{align}
where the ``quadrupole tensor'' is
\be\label{Qij}
Q_{ij}(a) \equiv -\frac{2}{D_+(a)}\int_{a_i}^a ds\, \mathcal{G}(a,s)\frac{\sigma_{ij}(s)\dot{D}_+(s)}{s^2H(s)}\,,
\ee
which is defined in terms of the function
\be
\mathcal{G}(a, s) = \frac{2s^3}{5\Omega_m^0}\frac{H(s)}{H_0}[D_+(a)D_-(s) - D_-(a)D_+(s)]\,,
\ee
and where we have chosen conditions such that $\delta_1(a_i)=\delta_1^\iso(a_i)$.

\begin{figure}
	\centering
	\includegraphics[scale=0.55]{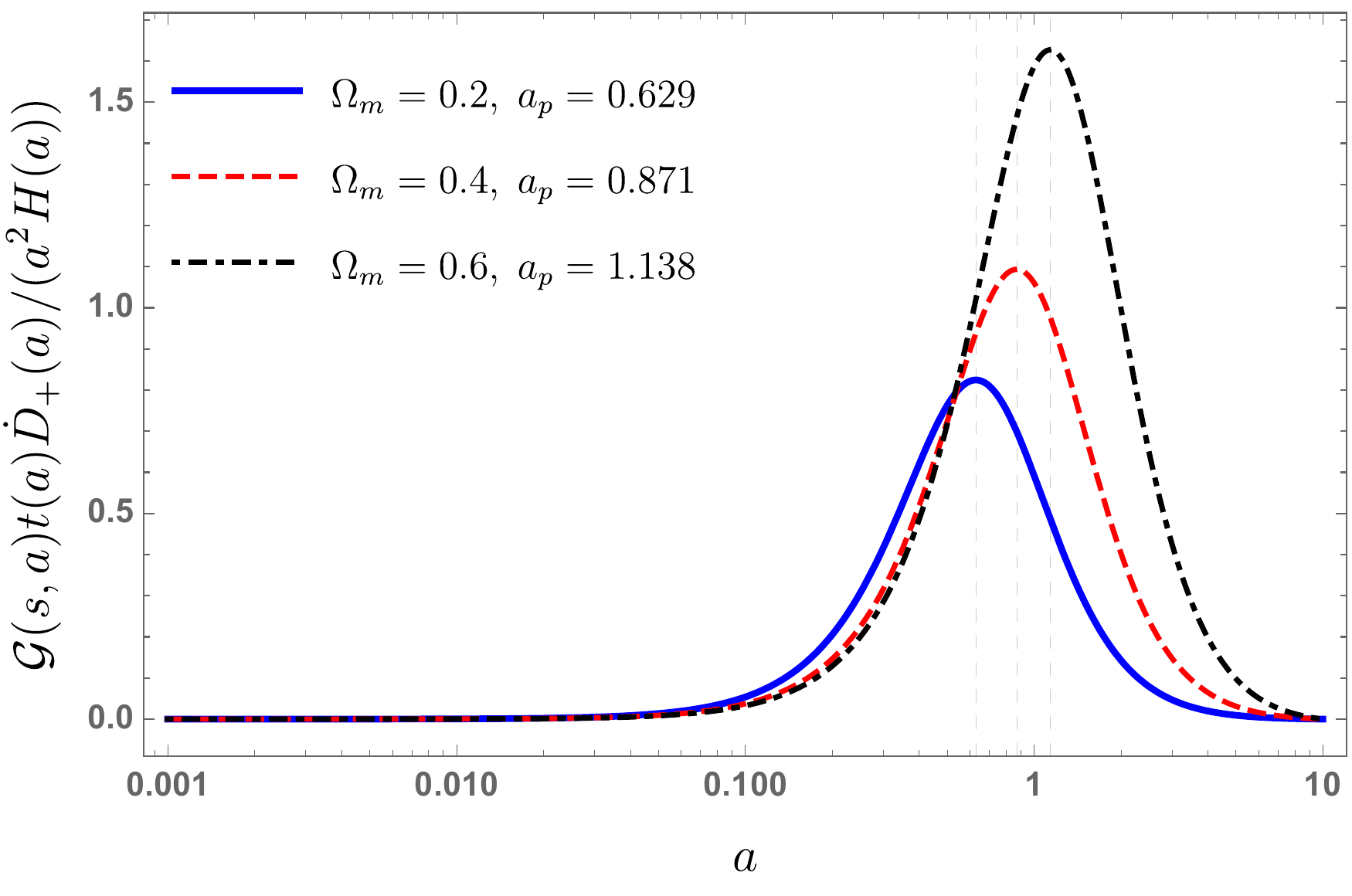}
	\caption{
	Time evolution of the integrand in the equation \eqref{Qij}. The vertical axis represents the kernel ${\cal G}(s,a)\frac{t(a)\dot{D}_+}{a^2H(a)}$ for a final scale factor $s=10$. We use a model with only matter and cosmological constant. The blue solid  line shows the evaluation of the kernel for $\Omega_m=0.2$,  the dashed red line for $\Omega_m=0.4$ and the black dot dashed line for $\Omega_m=0.6$. For the particular cases ploted here, the kernel peaks at $a_p=0.629$, $a_p=0.871$ and $a_p=1.138$ respectively. Note that we use an arbitrary normalization for the shear. }
	\label{fig:kernel_peak}
\end{figure}

Expression (\ref{delta-linear-lcdm}) is the solution we were after. Despite being model-dependent, it has some general and interesting properties. First, as noted above, it develops quadrupolar anisotropies through the shear which will propagate to the linear power spectrum. Second, it shows that a late-time phase of anisotropic expansion can only contribute to $\delta_1(a,\bfk)$ during a finite time window prior to the epoch $a$. In fact, since ${\cal G}(a,a)=0$, any model for $\sigma_{ij}$ which evolves from zero in the past --- say, at CMB decoupling --- to some non-zero value today implies that ${\cal G}(a,a_{\rm dec})\sigma_\parallel(a_{\rm dec},\bfk)\approx0$ at early times and ${\cal G}(a,a)\sigma_\parallel(a,\bfk)=0$ today. In other words, the kernel of integral in (\ref{Qij}) necessarily peaks at some redshift, though the precise value is model-dependent. 
We can nevertheless get a general idea of this behavior by assuming $\pi^i_{\ds j}$ to be a slowly varying function of $a$, in which case the growing mode of the shear (see \eqref{eq:shear}) can be written as $\sigma^i_{\ds j}=t(a)\pi^i_{\ds j}$. This is shown in Fig.~\ref{fig:kernel_peak}, which depicts the behaviour of the 
kernel of the integral in (\ref{Qij}).\\

\begin{figure}
	\centering
	\includegraphics[scale=0.52]{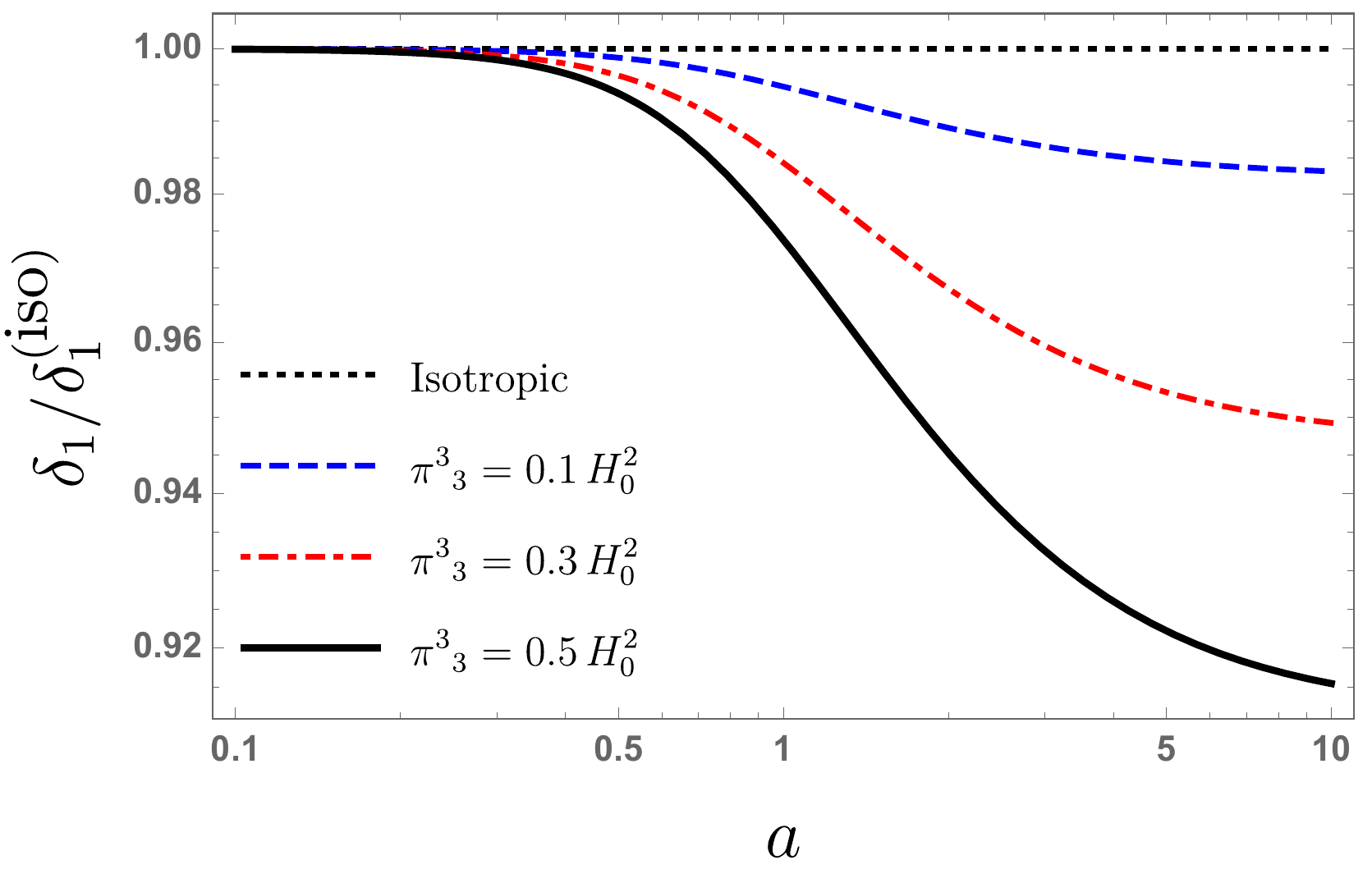}
	\includegraphics[scale=0.445]{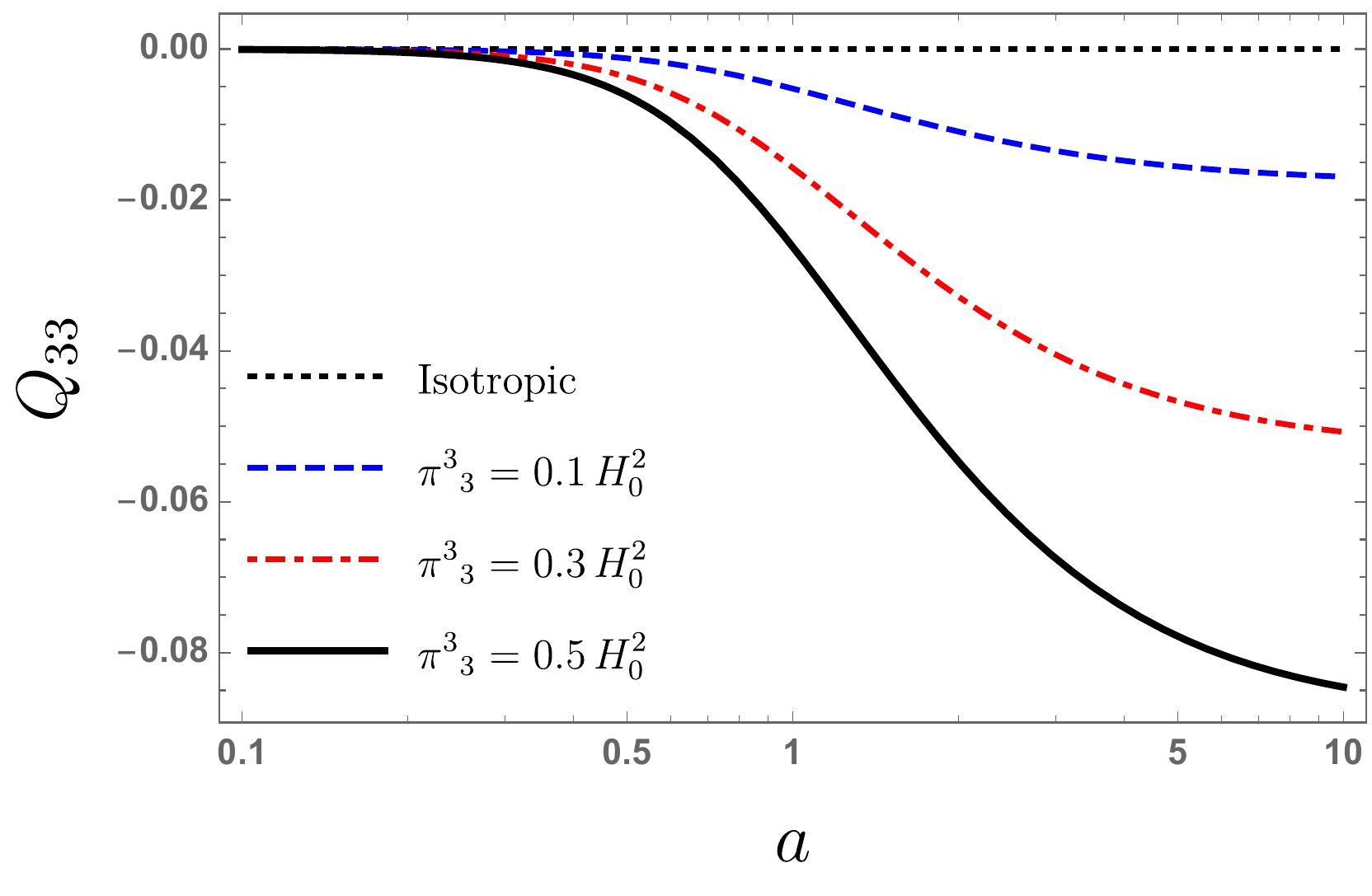}
	\includegraphics[scale=0.405]{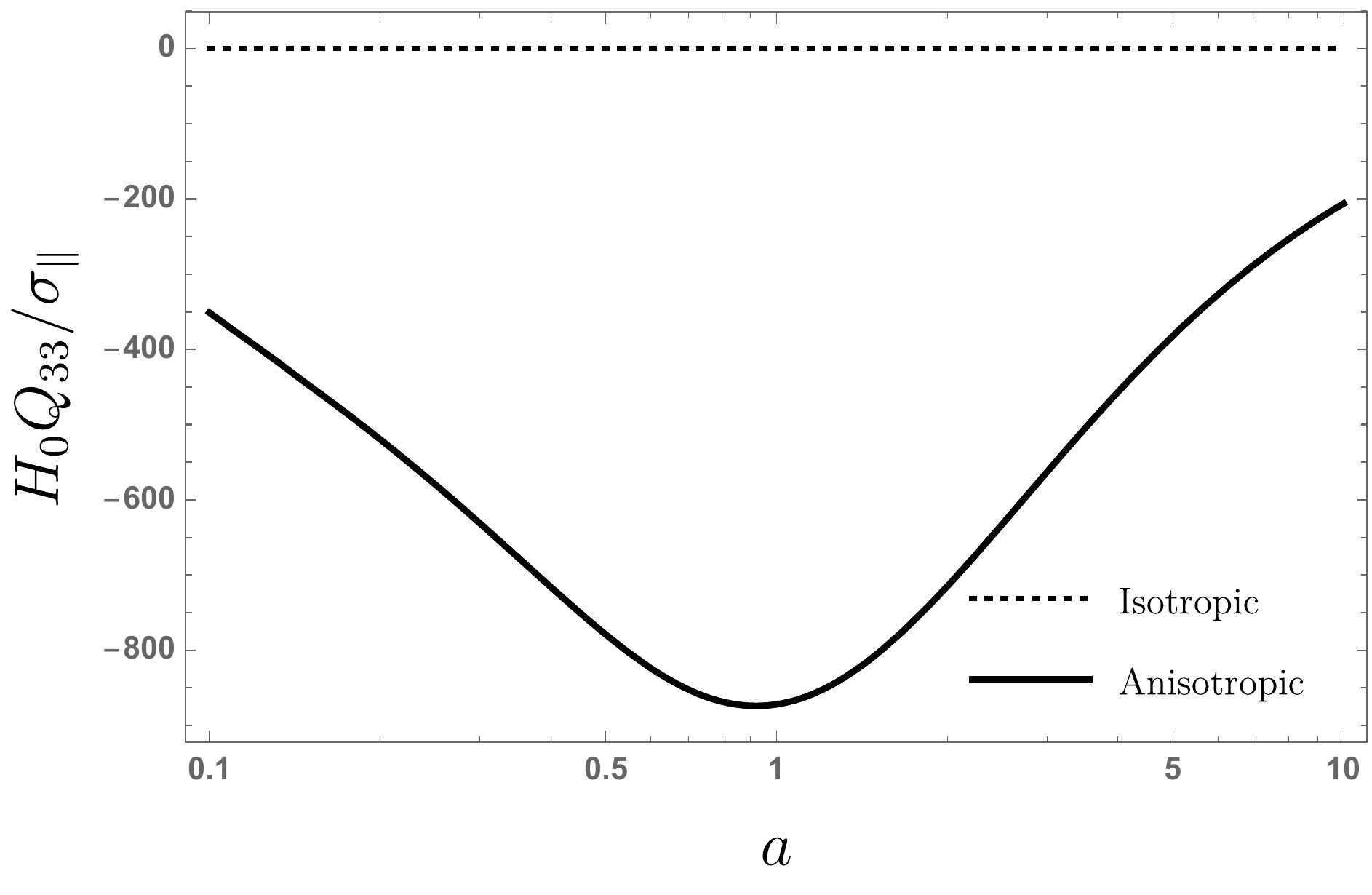}
	\caption{The effect of the shear on the linear matter density contrast. For definiteness we use a benchmark cosmology with only matter and cosmological constant with $\Omega_{m} = 0.23$. The unitary scale vector is  $\hat{\bfk} = (\sin \theta \cos \phi,\, \sin \theta \sin \phi,\, \cos \theta)$ and for the plot above we choose $\theta=0, \, \phi=0$. The upper panel shows the ratio of the anisotropic linear solution \eqref{delta-linear-lcdm} and the isotropic one for different values of the anisotropic stress $\pi^{3}{}_{3}=0.1 H_0^2,\, 0.3H_0^2, \, 0.5H_0^2$.  The lower left panel shows the  component $Q_{33}$ of the quadrupole tensor defined in (\ref{Qij}) and the lower right panel shows the ratio $H_0 Q_{33}/\sigma_{\parallel}$.}
	\label{fig:DplusQ}
\end{figure}

It is also instructive to have a general quantitative idea of the behavior of the linear density contrast in the presence of shear and compare it with the standard isotropic case.  To this end, in Fig.~\ref{fig:DplusQ} we plot the ratio of the anisotropic linear density contrast growth function and the isotropic density contrast $\delta_1/\delta_{1}^{(\rm iso)}$ for different values of the shear. In order to maximize the effect of the anisotropic contribution, we choose the direction of the unitary scale vectors $\hat{k}$ along the $z$-axis. We also show a plot of a particular component of the quadrupole term $Q_{ij}$ and the ratio $H_0Q_{ij}/\sigma_{\parallel}$ which is a dimensionless quantity independent of the amplitude of the anisotropy. For concreteness, in all the plots shown in Fig.~\ref{fig:DplusQ} we use as benchmark cosmology a flat $\Lambda$CDM model with only matter and cosmological constant with $\Omega_{m} = 0.23$ and a constant late time anisotropic stress tensor $\pi^{i}_{\ds j}$ in the equation  (\ref{eq:shear}).  

\subsection{Non-linear solutions}

We now wish to solve the fluid equations~\eqref{eomkdelta}-\eqref{eomktheta} to second order in perturbations.  In order to do this, we plug the linear equations of Section~\ref{sect:linear} in the non-linear terms of equations~\eqref{eomkdelta}-\eqref{eomktheta}. The relevant equations at second order are
\ba\label{eomkdelta2}
&&\delta_2' (\eta,{\bf{k}})  + \theta_2 (\eta,{\bf{k}}) =-\alpha[\theta_1, \delta_1]\,, \\
&& \theta_2' (\eta, {\bf{k}}) +  \left[{\cal{H}} +2\sigma_{\parallel}(\eta,\bfk)\right]\theta_2 (\eta,{\bf{k}}) + \frac{3}{2}{\cal{H}}^2\Omega_m \delta_2 (\eta, {\bf{k}}) = -\beta[\theta_1, \theta_1]\,, \label{eomktheta2}
\ea
where $X[\cdot,\cdot]$ denotes a convolution integral with the kernel $X$. Thus, for example
\begin{align}
\alpha[\theta_1, \delta_1] &\equiv \int [{\rm d}k]^2  (2\pi)^3\delta_D ({\bf k}-{\bf k}_{12})\alpha (\eta, {\bf k}_1, {\bf k}_2)\theta_1 (\eta,{\bf{k}}_1)\delta_1 (\eta, {\bf{k}}_2)\,, \\
\beta[\theta_1, \theta_1] &\equiv \int [{\rm d}k]^2  (2\pi)^3\delta_D ({\bf k}-{\bf k}_{12})\beta (\eta, {\bf k}_1, {\bf k}_2)\theta_1 (\eta,{\bf{k}}_1)\theta_1 (\eta, {\bf{k}}_2)\,.
\end{align}
For convenience, we also define\footnote{Note that, in this notation, $(\alpha[\theta_1,\delta_1])'\neq\alpha'[\theta_1,\delta_1]$.}
\be
\alpha'[\theta_1, \delta_1] \equiv \int [{\rm d}k]^2  (2\pi)^3\delta_D ({\bf k}-{\bf k}_{12})\alpha'(\eta, {\bf k}_1, {\bf k}_2)\theta_1 (\eta,{\bf{k}}_1)\delta_1 (\eta, {\bf{k}}_2)\,.
\ee

Equations~\eqref{eomkdelta2}-\eqref{eomktheta2} can now be combined in order to obtain an equation for $\delta_2$. Using the linear equations to replace $\theta'_1$ in terms of $\delta_1$ and its derivatives, we arrive at
\begin{align}\label{eq:delta2eta}
\delta_2''(\eta, {\bf k}) + [\mathcal{H} + 2\sigma_\parallel(\eta, {\bf k})]\delta_2'(\eta, {\bf k}) & - \frac{3}{2}\mathcal{H}^2\Omega_m \delta_2(\eta, {\bf k}) \nonumber\\ 
=\; &  \frac{3}{2}\mathcal{H}^2\Omega_m \alpha[\delta_1, \delta_1] + \beta[\theta_1,\theta_1] - \alpha[\theta_1, \delta'_1] \nonumber \\
&-2\sigma_\parallel\alpha[\theta_1, \delta_1] + 2 \alpha[\sigma_\parallel \theta_1, \delta_1] - \alpha'[\theta_1, \delta_1]\,.
\end{align}
Finally, we exchange derivatives with respect to time for derivatives with respect to the scale factor, to obtain
\begin{align}\label{eq:delta2a}
\ddot{\delta}_2(a, {\bf k}) + \left(\frac{\dot{H}}{H} + \frac{3}{a}\right)\dot{\delta}_2(a, {\bf k}) 
& - \frac{3}{2}\frac{\Omega^0_m}{a^5}\left(\frac{H_0}{H}\right)^2 \delta_2(a, {\bf k}) + \frac{2\sigma_\parallel}{a^2 H}\dot{\delta}_2(a, {\bf k}) \nonumber\\ 
=\; & \frac{3}{2}\frac{\Omega^0_m}{a^5}\left(\frac{H_0}{H}\right)^2 \alpha[\delta_1, \delta_1] + \beta[\dot{\delta}_1,\dot{\delta}_1] + \alpha[\dot{\delta}_1, \dot{\delta}_1] \nonumber \\
& +\frac{2\sigma_\parallel}{a^2 H}\alpha[\dot{\delta}_1, \delta_1] - \frac{2}{a^2H} \alpha[\sigma_\parallel \dot{\delta}_1, \delta_1] + \dot{\alpha}[\dot{\delta}_1, \delta_1]\,.
\end{align}

Before we proceed to solve the above equation, it is convenient to split the density contrast, at each order, into an isotropic piece and an anisotropic correction. We thus write
\be
\begin{aligned}
\delta_1 & = \delta_1^{\iso} + \Delta_1\,,\\
\delta_2 & = \delta_2^{\iso} + \Delta_2\,,
\end{aligned}
\label{eq:Delta2def}
\ee
where $\Delta_1$ can be read off from Eq.~\eqref{delta-linear-lcdm}, and $\Delta_2$ is what we are after. Next, we solve for the isotropic piece $\delta_2^{\iso}$, and then compute the correction due to the presence of a small anisotropy $\Delta_2$. For this purpose, we also split the kernels (\ref{alpha-beta}) into an isotropic piece and an anisotropic correction:
\begin{align}\label{alphabeta}
\begin{split}
\alpha(\eta, {\bf k}_1, {\bf k}_2) & = \alpha^{\iso}({\bf k}_1, {\bf k}_2) + \Delta\alpha(\eta, {\bf k}_1, {\bf k}_2)\,, \\
\beta(\eta, {\bf k}_1, {\bf k}_2) & = \beta^{\iso}({\bf k}_1, {\bf k}_2) + \Delta\beta(\eta, {\bf k}_1, {\bf k}_2)\,.
\end{split}
\end{align}
At linear order in the shear, the background metric (\ref{misnermetric}) can be parametrized as
\be
\gamma_{ij} = [e^{2\lambda}]_{ij} \approx \delta_{ij} + 2\lambda_{ij}\,.
\ee
Defining $\lambda_\parallel(\eta,\bfk)\equiv \lambda_{ij}(\eta)\hat{k}^i\hat{k}^j$, in analogy to (\ref{sigmapar}), this allows us to rewrite the correction to the kernels as
\begin{align}
\Delta\alpha(\eta, \bfk_1, \bfk_2) & = 2\lambda_{ij}(\eta)\frac{ k^i_{12} k^j_1}{k_1^2} - 2\lambda_\parallel(\eta, \bfk_1)\alpha^{\iso}(\bfk_1, \bfk_2)\,,\label{deltaalpha}\\
\Delta\beta(\eta,\bfk_1,\bfk_2) & = \lambda_{ij}(\eta)\frac{k_1^i k_2^j (\bfk_{12})^2}{k_1^2 k_2^2} + 2(\lambda_\parallel(\eta, \bfk_{12}) - \lambda_\parallel(\eta, {\bf k}_1) - \lambda_\parallel(\eta, {\bf k}_2))\beta^{\iso}(\bfk_1, \bfk_2)\,.\label{deltabeta}
\end{align}
We will need this in order to solve for $\Delta_2$.
\subsubsection{Review of the isotropic case}
As a consistency check, let us recover the standard (nonlinear) isotropic solution. In this case, equation (\ref{eq:delta2a}) takes the form
\begin{multline}
\label{eq:delta2iso}
\ddot{\delta}^{\iso}_2(a, {\bf k}) + \left(\frac{\dot{H}}{H} + \frac{3}{a}\right)\dot{\delta}^{\iso}_2(a, {\bf k}) - \frac{3}{2}\frac{\Omega^0_m}{a^5}\left(\frac{H_0}{H}\right)^2 \delta^{\iso}_2(a, {\bf k}) \\ = \frac{3}{2}\frac{\Omega^0_m}{a^5}\left(\frac{H_0}{H}\right)^2 \alpha^{\iso}[\delta^{\iso}_1, \delta^{\iso}_1] + \alpha^{\iso}[\dot{\delta}^{\iso}_1,\dot{\delta}^{\iso}_1] + \beta^{\iso}[\dot{\delta}^{\iso}_1,\dot{\delta}^{\iso}_1]\,.
\end{multline}
At second order in perturbations, we only consider the nonhomogeneous solution of the equation above. This can again be obtained by using the Wronskian of the homogeneous (first order) solutions, thus finding
\be\label{sol-delta2-iso}
\delta_2^\iso(a, {\bf k}) =  \int \frac{d^3 q}{(2\pi)^3}F_2^{\iso}(a, {\bf q}, {\bf k} - {\bf q}) \delta_l({\bf q})\delta_l( {\bf k} - {\bf q})\,,
\ee
where the isotropic second order kernel reads
\begin{multline}\label{F2iso}
F_2^{\iso}(a, {\bf k}_1, {\bf k}_2) = \int^a_{a_i} ds\ \mathcal{G}(a,s)\bigg\{\dot{D}_+^2(s)\beta^{\iso}({\bf k}_1,{\bf k}_2) \\ + \left[\frac{3}{4}\frac{\Omega_m^0}{s^5}\left(\frac{H_0}{H(s)}\right)^2 D_+^2(s) + \frac{1}{2}\dot{D}_+^2(s)\right]\left(\alpha^{\iso}({\bf k}_1,{\bf k}_2) + \alpha^{\iso}({\bf k}_2,{\bf k}_1)\right)\bigg\}\,.
\end{multline}

Expression (\ref{sol-delta2-iso}) gives the most general solution at second order in a $\Lambda$CDM universe. However, as one can see from $F^\iso_2$, its space and time dependencies are not separable, and thus this solution is not appropriate to write higher order solutions recursively. While we shall not be interested in writing solutions higher than second order in this work, it is worth mentioning that, as is well known~\cite{Scoccimarro:1997st,Bernardeau:2001qr}, one can obtain separable solutions by approximating the linear growth rate as
\be\label{separability}
f(\Omega_m,\Omega_\Lambda) = \frac{d\log D_+}{d\log a} \approx \Omega_m^{1/2}\,,
\ee
which is equivalent to using $\dot{D}_{+}=(\Omega_m^0/a^5)^{1/2}(D_{+}H_0/H)$. Furthermore, one can check that the decaying mode goes as $D_{-} = D_{+}^{-3/2}$ under this approximation (see Appendix \ref{sec:separability}). In this case $\mathcal{G}(a,s) = \frac{2s}{5f}(D_+(a) D^{-1}_+(s) - D_+^{-3/2}(a)D_+^{3/2}(s))$. Using all of this, we find the standard (separable) kernel:
\begin{align}\label{F2isoApprox}
F_2^{\iso}(a, {\bf k}_1, {\bf k}_2) = D_{+}^2(a) \bigg\{\frac{5}{14} [\alpha^{\iso}({\bf k}_1,{\bf k}_2) + \alpha^{\iso}({\bf k}_2,{\bf k}_1)] + \frac{2}{7}\beta^{\iso}({\bf k}_1,{\bf k}_2)\bigg\}\,.
\end{align}
However, we shall not use approximation (\ref{separability}) in our results, since the presence of a background shear precludes any separability. We also stress that, in what follows, we shall be following the standard practice of dropping terms proportional to $D_+(a_i)$, since these terms grow slower than the ones we kept.
\subsubsection{Anisotropic corrections}
We now finally solve for $\Delta_2$. For that we plug Eq.~\eqref{eq:Delta2def} into Eq.~\eqref{eq:delta2a}, using also Eq.~\eqref{eq:delta2iso}, to find an equation for $\Delta_2$. The resulting equation is somewhat long, but can be easily solved and simplified. It is given by
\begin{align}\label{eq:Delta2a}
\ddot{\Delta}_2(a, {\bf k}) + & \left(\frac{\dot{H}}{H} + \frac{3}{a}\right)\dot{\Delta}_2(a, {\bf k}) - \frac{3}{2}\frac{\Omega^0_m}{a^5}\left(\frac{H_0}{H}\right)^2 \Delta_2(a, {\bf k}) \nonumber\\ 
= & - \frac{2\sigma_\parallel}{a^2 H}\dot{\delta}^{\iso}_2 + \Delta\beta[\dot{\delta}^{\iso}_1,\dot{\delta}^{\iso}_1] + 2\beta^{\iso}[\dot{\Delta}_1,\dot{\delta}^{\iso}_1] \nonumber\\
& + \frac{3}{2}\frac{\Omega^0_m}{a^5}\left(\frac{H_0}{H}\right)^2 (\Delta\alpha[\delta^{\iso}_1, \delta^{\iso}_1]  + \alpha^{\iso}[\Delta_1, \delta^{\iso}_1]+\alpha^{\iso}[\delta^{\iso}_1,\Delta_1]) \nonumber\\
& + \Delta\alpha[\dot{\delta}^{\iso}_1, \dot{\delta}^{\iso}_1] + \alpha^{\iso}[\dot{\Delta}_1, \dot{\delta}^{\iso}_1]+ \alpha^{\iso}[\dot{\delta}^{\iso}_1,\dot{\Delta}_1] + \Delta\dot{\alpha}[\dot{\delta}^{\iso}_1, \delta^{\iso}_1] \nonumber \\
&+\frac{2\sigma_\parallel}{a^2 H}\alpha^{\iso}[\dot{\delta}^{\iso}_1, \delta^{\iso}_1] - \frac{2}{a^2H} \alpha^{\iso}[\sigma_\parallel \dot{\delta}^{\iso}_1, \delta^{\iso}_1]\,.
\end{align}
Note that, despite being similar, the last two terms on the right-hand side are not the same. 

The (nonhomogeneous) solution to this equation can be obtained using the Wronskian, as before. The solution can be grouped into different time-dependent quadrupolar terms, and the final result can be written in a compact form:
\be\label{Delta2final}
\Delta_2(a, {\bf k}) =  \int \frac{d^3 q}{(2\pi)^3} F^{\sigma}_2(a,  {\bf q},{\bf k} - {\bf q})\delta_l({\bf q})\delta_l({\bf k} - {\bf q})
\ee
where
\begin{align}
F^\sigma_2(a, {\bf k}_1, {\bf k}_2) &= A_{ij}(a)\hat{k}_{12}^i\hat{k}_{12}^j\big(\alpha^{\iso}({\bf k}_1, {\bf k}_2) + \alpha^{\iso}({\bf k}_2, {\bf k}_1)\big) + B_{ij}(a)\hat{k}_{12}^i\hat{k}_{12}^j\beta^{\iso}({\bf k}_1, {\bf k}_2) \nonumber \\
&\phantom{=}+C_{ij}(a)\left(\hat{k}_1^i\hat{k}_1^j\alpha^{\iso}({\bf k}_1, {\bf k}_2) + \hat{k}_2^i\hat{k}_2^j\alpha^{\iso}({\bf k}_2, {\bf k}_1)\right)+E_{ij}(a)\left(\frac{k_{12}^ik_1^j}{k_1^2} + \frac{k_{12}^i k_2^j}{k_2^2}\right) \nonumber \\
&\phantom{=}+H_{ij}(a)(\hat{k}_1^i\hat{k}_1^j + \hat{k}_2^i\hat{k}_2^j)\beta^{\iso}({\bf k}_1, {\bf k}_2) + K_{ij}(a)\hat{k}_1^i\hat{k}_2^j\frac{({\bf k}_{12})^2}{k_1k_2}\,.\label{eq:F2sigma}
\end{align}
The explicit expressions for the tensors $A$, $B$, $C$, $E$, $H$, $K$ can be found in Appendix~\ref{app:quadrupolar_coefs}.

\section{Correlation functions}\label{sec:correlations}
We now give the formal expressions for the (tree level) two- and three-point correlation functions for the dark matter density contrast in a late-time Bianchi I universe. For generality, we write the correlations at different times. We will need them in order to comment on the ``Galilean symmetries'' of our solutions.

\subsection{Power spectrum and bispectrum}
We start with the two-point function. Since our solution \eqref{delta-linear-lcdm} splits into a stochastic Gaussian variable multiplied by a time-dependent amplitude, the two-point correlation for the density contrast at different times is readily obtained:
\be\label{powers}
\langle\delta(a_1,\mathbf{k})\delta(a_2,\mathbf{q})\rangle = (2\pi)^3P_L(a_1,a_2,\bfk)\delta_D(\mathbf{k}+\mathbf{q})\,,
\ee
where
\be
P_L(a_1,a_2,\bfk) = P_L^\iso(a_1, a_2,k)\left(1+Q_{ij}(a_1)\hat{k}^i\hat{k}^j+Q_{ij}(a_2)\hat{k}^i\hat{k}^j\right)\,,
\ee
and $P_L^\iso(a,k)$ is the $\Lambda$CDM linear power spectrum (see, \emph{e.g.}, \cite{dodelson2020modern}).
As emphasized before, the effect of the background anisotropy is to induce quadrupolar corrections in the power spectrum, which is here represented by the $Q_{ij}\hat{k}^i\hat{k}^j$ terms. Given a model for anisotropic dark-energy, these corrections can be directly computed, in a spherical basis, as ${\cal Q}_{2m} = \int d^2\hat{\bfk}Q_{ij}\hat{k}^i\hat{k}^jY^*_{2m}(\hat{\bfk})$. 
In practice, however, the extraction of some signal from the data is complicated by the fact that cosmological surveys probe the power spectrum in redshift space, which also contains a quadrupolar correction due to redshift-space distortions \cite{Kaiser:1987qv,Hamilton:1997zq}. Thus, in order to constrain the anisotropy of dark energy, it is necessary to disentangle these signatures from the measured quadrupole \cite{Yamamoto:2005dz,BOSS:2013uda}. This is not a trivial task, and requires in particular that we formulate redshift distortions and linear biasing in an anisotropic background. We leave these questions to a future work.

The three-point correlation function is computed from the total contrast, $\delta=\delta_1+\delta_2$, keeping only second order terms in $\delta_1$ and linear terms in $\Delta_1$. The (connected) function at different times is of the form
\be
\langle\delta(a_1, \bfk_1)\delta(a_2, \bfk_2)\delta(a_3, \bfk_3)\rangle = (2\pi)^3 B(\{a\}, \bfk_1, \bfk_2, \bfk_3)\,,
\ee
where $\{a\}=(a_1,a_2,a_3)$, and the bispectrum is given by
\be
\label{eq:isobispectrum}
B(\{a\}), \bfk_1, \bfk_2, \bfk_3) = 2F_2(a_3, \bfk_1, \bfk_2)P_L(a_1,a_3, \bfk_1)P_L(a_2,a_3, \bfk_2) + \mathrm{cycl.}\,.
\ee
Here, we take $F_2 = F_2^{\iso} + F_2^{\sigma}$. Similarly, by writing $B = B^{\iso} + B^\sigma$ we obtain
\begin{multline}\label{powerb}
B^\sigma(\{a\}, \bfk_1, \bfk_2, \bfk_3) = 2\bigg[F_2^{\iso}(a_3, \bfk_1, \bfk_2)(Q_{ij}(a_1)\hat{k}_1^i\hat{k}_1^j + Q_{ij}(a_2)\hat{k}_2^i\hat{k}_2^j) \\
+ F_2^\sigma(a_3, \bfk_1, \bfk_2)\bigg]P_L^{\iso}(a_1,a_3, k_1)P_L^{\iso}(a_2,a_3, k_2)+ \mathrm{cycl.}\,,
\end{multline}
which is our final result.
\subsection{Comment on a ``Galilean symmetry''}
Apart from rotational invariance, the Bianchi I background is less symmetric than the usual FLRW background in other, non-obvious, ways. The latter has a ``Galilean symmetry'' \cite{Kehagias:2013yd, Peloso:2013zw}. That is, one can find a coordinate and field transformation which leave the continuity, Poisson, and Euler equations  invariant, and at the same time sets the long-wavelength velocity to zero. In other words, the effect of a long-wavelength mode is equivalent to a transformation which leaves the equations invariant.\footnote{In this sense, it is more akin to the equivalence principle than a Galilean symmetry: One can find a frame which falls freely in the long-wavelength gravitational potential.} This was shown to come from the Newtonian limit of large gauge transformations in general relativity~\cite{Creminelli:2013mca}. These gauge transformations at infinite wavelength which can be continuously deformed to finite wavelength are called ``adiabatic modes''~\cite{Weinberg:2003sw}. This ``Galilean symmetry'' can be used, for example, to fix the leading terms of a zero-momentum limit of an $n$-point function in terms of an $(n-1)$-point function, giving a ``consistency relation'' between them. In the case of the bispectrum $B(\{a\}, {\bfk_1}, {\bfk_2}, {\bf{q}})$, one can use this to fix the
$\mathcal{O}(k/q)$ term in the $q/k \ll 1$ expansion. At equal times, this $k/q$ term in the expansion is always zero due to translational invariance, parity, and the exchange symmetry $k_1 \leftrightarrow k_2$. At different times, it is fixed by ``Galilean invariance'' to be
\be
\lim_{q \rightarrow 0}B^\iso(\{a\}, {\bfk_1}, {\bfk_2}, {\bf{q}}) =-
\frac{{\bf q}\cdot{\bf k}
}{q^2} \frac{(D_+(a_1)  - D_+(a_2))}{D_+(a_3)}   P(a_3, a_3,q)P(a_1, a_2, k) + \mathcal{O}\left(\frac{q^0}{k^0}\right)\,,
\label{eq:consitency}
\ee
where we kept only leading terms and ${\bf k} = {\bfk}_1 \simeq -{\bfk}_2$. One can check that the tree-level isotropic part of the bispectrum~\eqref{eq:isobispectrum} satisfies this relation. This was also used recently to fix the behaviour of non-linear kernels in perturbation theory using only these ``symmetries'' in a bootstrap  approach~\cite{DAmico:2021rdb}.

As a check of whether Bianchi I has the same ``symmetries'' let us take the squeezed limit of the quadratic kernel~\eqref{eq:F2sigma}. We get
\begin{multline}
\lim_{q \rightarrow 0} F_2^\sigma({\bf q}, {\bf k}) = \left(A_{ij} + \frac{1}{2}B_{ij} + \frac{1}{2}H_{ij}\right) \hat{k}^i \hat{k}^j \frac{{\bf q}\cdot{\bf k}}{q^2}  \\ + \left(C_{ij} + \frac{1}{2}H_{ij}\right) \hat{q}^i \hat{q}^j \frac{{\bf q}\cdot{\bf k}}{q^2} + (K_{ij} + E_{ij})\frac{q^i k^j}{q^2} + \mathcal{O}\left(\frac{q^0}{k^0}\right)\,.
\label{eq:F2squeezed}
\end{multline}
Using this in equation~\eqref{powerb}, one can check that the consistency relation at different times, Eq.~\eqref{eq:consitency}, is not satisfied.\footnote{At equal times they are still zero since translational invariance, parity, and exchange symmetry still hold.}

This can be somewhat surprising, since Eqs.~\eqref{eq:continuity_nr}, \eqref{eq:full_euler}, and \eqref{eq:Poisson} still seem to have a ``Galilean symmetry''. Let us discuss the matter dominated case for simplicity. One can write a simple generalization of the ``Galilean'' transformations which leave the equations invariant
\begin{align}
\begin{split}
{\bf x} &\rightarrow \tilde{{\bf x}} = {\bf x} + {\bf b} \eta^2/2\,,\\
\delta({\bf x}, \eta) &\rightarrow \tilde{\delta}(\tilde{\bf x}, \eta) = \delta({\bf x}, \eta)\,,\\
{\bf v}({\bf x}, \eta) &\rightarrow \tilde{\bf v}(\tilde{\bf x}, \eta) = {\bf v}({\bf x}, \eta) - {\bf b}\eta\,,\\
\phi({\bf x}, \eta) &\rightarrow \tilde{\phi}(\tilde{\bf x}, \eta) = \phi({\bf x}, \eta) + 3{\bf b}\cdot
{\bf x} + 2\eta \sigma_{ij}x^i b^j\,,
\end{split}
\end{align}
where $\mathbf{b}$ is a constant vector. The coordinate transformation and the transformation of the density contrast are the same as before. One would then naively expect the same consistency relation to hold.
However, these transformations are \emph{not} adiabatic modes. That is, they  cannot be used to set the long-wavelength velocity to zero. Indeed, consider the long-wavelength velocity from our linear solution
\be\label{vzero}
{\bf v}(\eta, {\bf q}) = {\bf v}(\eta_i,{\bf q})\ \frac{\eta}{\eta_i}\left[1 - \frac{4}{5}\int_{\eta_i}^\eta\left(\frac{\eta^5-y^5}{\eta^5}\right)\sigma_\parallel(y,{\bf q} ) dy - \frac{2}{5} \eta\frac{d}{d\eta}\int_{\eta_i}^\eta\left(\frac{\eta^5-y^5}{\eta^5}\right)\sigma_\parallel(y,{\bf q}) dy\right]\,.
\ee
Due to the non-trivial time dependence of the terms proportional to $\sigma_\parallel$, this transformation cannot be used to compute the effect of a long mode on short-wavelength perturbations. Therefore, we don't obtain a consistency relation from this invariance. Note, however, that equation \eqref{eq:F2squeezed} vanishes after averaging over the angle between the long and short modes. This hints at the existence of an angle-averaged adiabatic mode.

\section{Conclusions and discussion}\label{sec:conclusions}

In this work we used Eulerian perturbation theory to describe the evolution of the large scale structure in a late-time anisotropic universe. We assumed that the anisotropy is sourced by a small stress component in the dark energy fluid. 
We took the Newtonian limit of the Einstein and stress-energy conservation equations in a Bianchi I background, and worked in the limit of small anisotropies. We thus obtained equations~\eqref{eq:delta}-\eqref{eq:theta}, which give the Eulerian description for the evolution of the dark matter density contrast and velocity field in this setup. 
We then solved these equations perturbatively, obtaining expressions for the linear density contrast~\eqref{Qij} and the second order kernel~\eqref{eq:F2sigma} in the presence of anisotropies. 
This allowed us to compute the power spectrum~\eqref{powers} and bispectrum~\eqref{powerb} of the dark matter density contrast. Quite generally, we found that the contribution of anisotropies  to the linear growth of dark matter will peak at some non-zero redshift, although the exact redshift value is model dependent. 
This feature, together with the complex momentum structure of the quadrupolar terms in Eq.~\eqref{eq:F2sigma}, can in principle be used to disentangle these effects from the quadrupolar corrections induced by redshift space distortions using, \emph{e.g.}, the estimator of \cite{Scoccimarro:2015bla}. 

Regarding our second-order solution, note that some of the terms in equation~\eqref{eq:F2sigma} depend on $\lambda_{ij}$ today, 
rather than on its time derivatives, $\sigma_{ij}$. Thus, they seem to contribute to the bispectrum when $\lambda_{ij}$ is constant. This is not surprising and, as we argue in Appendix~\ref{app:quadrupolar_coefs}, it happens because a constant $\lambda_{ij}$ is equivalent to an anisotropic coordinate rescaling of the FLRW metric $x^i \rightarrow [e^{2\lambda}]_{ij} x^j$. Therefore, these terms should have no observable effects, and are expected to disappear once we compute the power spectrum and bispectrum in redshift space. We leave this task for future work.

We also discussed the role of Galilean symmetries~\cite{Kehagias:2013yd, Peloso:2013zw} in an anisotropic universe. We showed that the usual consistency relation does not hold, meaning that the adiabatic modes of a Bianchi I universe would have to be very different from the isotropic case.  We conclude that one can't use a simple extension of the Galilean symmetries in order to carry out a bootstrap approach~\cite{DAmico:2021rdb} in this case. Thus, theories with a Bianchi I background can in principle have more distinct observational signatures than isotropic modified gravity and dark energy theories. It is possible, however, that less trivial extensions to the usual Galilean symmetries can be used to set the long-wavelength velocity to zero, and we leave the question of (anisotropic) adiabatic modes for a future investigation.

Equations~\eqref{eq:delta}-\eqref{eq:theta} are in principle valid at all orders in perturbation theory. However, they describe the evolution of a perfect fluid. Such description of the dark matter perturbations is approximately valid at large scales, but receives corrections due to the effect of coarse-graining over the small scales \cite{Baumann:2010tm}. This can be accounted for by introducing small counter-terms in the equations, that guarantee that loop corrections are well-behaved \cite{Carrasco:2012cv}. The new anisotropic counter-terms will be further suppressed by the smallness of the anisotropy, and we expect them to have a negligible effect.

The Bianchi I metric can also be used to describe the local effects of a very long wavelength tensor mode $\tau_{ij}$ \cite{Pontzen:2010eg,Pereira:2019mpp}. Indeed, in the limit in which the wavelength of the tensor mode goes to zero, such a perturbation will locally be described by the metric~\eqref{BI-metric}, where $\lambda_{ij} = \tau_{ij}$. If $\tau'_{ij} = 0$ there is then no locally observable effect. On the other hand, the time evolution of tensor modes inside the horizon can leave a small imprint on the large scale structure~\cite{Schmidt:2013gwa}, and our method can also be used to compute this effect at linear order in $\tau_{ij}$ and arbitrary order in $\delta$.

We conclude by noting that, while we have focused on computing the evolution of the dark matter density contrast, the main observable in near future surveys will be galaxy number counts. They are expected to be related since galaxies form in potential wells sourced mainly by the dark matter. This relation is called ``bias'', see~\cite{Desjacques:2016bnm} for a recent review. In the anisotropic case, new anisotropic bias operators will need to be included at each order, and it would be interesting to describe these operators in detail.

\section*{Acknowledgments}
JPBA and CAVT acknowledge partial financial support from the Patrimonio Autónomo - Fondo Nacional de Financiamiento para la Ciencia, la Tecnología y la Innovación Francisco José de Caldas (MINCIENCIAS - COLOMBIA) Grant No. 110685269447 RC-80740-465-2020, project 69723. JM acknowledges financial support from  Ministerio de Ciencia, Tecnología e Innovación -- Colombia. JN is supported by FONDECYT grant 1211545, ``Measuring the Field Spectrum of the Early Universe''. TSP is supported by Brazilian funding agencies CAPES (Coordenação de Aperfeiçoamento de Pessoal de Nível Superior) and CNPq (Conselho Nacional de Desenvolvimento Científico e Tecnológico), Grants No. 438689/2018-6 and  311527/2018-3. 

\appendix

\section{Christoffel symbols}\label{christoffels}
We present here the Christoffel symbols which were used to compute the perturbed continuity and Euler equations in the main text. Recall that a prime means derivative with respect to conformal time, and that time derivatives do not commute with index manipulations. To order ${\cal O}(\epsilon^{1/2})$, they are given by:
\begin{align}
\Gamma_{00}^{0} & ={\cal H}+{\cal O}(\epsilon)\,,\qquad\\
\Gamma_{0j}^{i} & ={\cal H}\delta_{j}^{i}+\sigma_{j}^{i}+{\cal O}(\epsilon)\,,\qquad\\
\Gamma_{ij}^{0} & ={\cal H}\gamma_{ij}+\sigma_{ij}+{\cal O}(\epsilon)\,,\qquad\\
\Gamma_{00}^{i} & =\partial^{i}\phi+{\cal O}(\epsilon^{3/2})\,,\qquad\\
\Gamma_{0i}^{0} & =\partial_{i}\phi+{\cal O}(\epsilon^{3/2})\,,\qquad\\
\Gamma_{jk}^{i} & =\frac{1}{2}\gamma^{il}\left(\partial_{j}h_{kl}+\partial_{k}h_{jl}-\partial_{l}h_{jk}\right)+{\cal O}(\epsilon^{3/2})\,.\qquad
\end{align}
Some particularly useful expressions are:
\begin{align}
\Gamma_{\mu0}^{\mu} & = 4{\cal H}+{\cal O}(\epsilon)\,,\\
\Gamma_{\mu i}^{\mu} & =\partial_{i}\phi-3\partial_{i}\psi+{\cal O}(\epsilon^{3/2})\,.
\end{align}

\section{Transverse velocity}\label{transversev}
Here we show that the non-decaying part of the transverse velocity is sourced linearly by the background shear tensor, and thus contributes as a subleading term in Euler equation. We start by recalling that, in Fourier space, the velocity is decomposed in its longitudinal and transverse parts as
\be
v^i(\bfk) = k^iv(\bfk)+v^i_\perp(\bfk)\,,\qquad k_iv^i_\perp(\bfk)=0\,.
\ee
Being orthogonal to $k^i$, the transverse velocity $v^i_\perp$ lives in a 2D plane spanned by unit-vectors $e^b_i$ ($b=1,2$), implicitly dependent on $k^i$, such that
\be
v^b_\perp(\bfk) \equiv e^b_iv^i(\bfk) = e^b_iv^i_\perp(\bfk)\,.
\ee
We can thus find the dynamical equation for the transverse velocity by projecting Euler's equation along $e^b_i$. In a Bianchi I universe, $e^b_i$ evolves in time according to~\cite{Pereira:2007yy}
\be
(e^b_i)' = -\sigma_{jl}e^j_be^l_ce^c_i + 2\sigma_{ij}e^j_b\,,
\ee
so that, at linear order, equation (\ref{eq:full_euler}) gives
\be
(v^b_\perp)'+{\cal H}v^b_\perp = -v^c_\perp\sigma_{jl}e^j_be^l_c\,.
\ee
To linear order in $\sigma_{ij}$, the solution to this equation is
\be\label{vperp}
v^b_\perp = \frac{v^b_0}{a}-\frac{v^c_0}{a}\int^\eta \sigma_{jl}e^j_ce^l_b d\eta' + {\cal O}(\sigma^2)\,.
\ee
Thus, the homogeneous solution decays, while de non-homogeneous is proportional to the shear. Since this solution enters Eq.~\eqref{eq:full_euler} through a term containing $\sigma_{ij}$, it does not contribute at linear order in anisotropies. One can similarly check that the non-linear terms in (\ref{vperp}) remain zero if not sourced initially.

\section{Separable solutions in the isotropic case}\label{sec:separability}

In this Appendix, we detail the derivation of the linear solutions for the fluid equations under the separability condition $f\approx \Omega_m^{1/2}$. As is well known, the fluid equations in an Einstein-de Sitter (EdS) universe can be exactly separated, at all orders in perturbations, into time and space. In a $\Lambda$CDM universe, such separation is only possible if $f \approx \Omega_m^{1/2}$, in which case the spatial part of the solutions is exactly the one of the EdS case~\cite{Scoccimarro:1997st,Bernardeau:2001qr}, thus greatly simplify calculations of the kernels. In our case, since we derived the nonlinear solution from the Wronskian of the (linear) homogeneous solutions, it is not obvious how this approximation is implemented, and we thus quickly review it here.

Let us begin by rewriting the linear equation for $\delta^{\iso}$, Eq.~\eqref{eomkdelta_linear} with $\sigma_\parallel = 0$, in terms of the variable $D \equiv D_+(\eta)$. Since this is the growing mode of the linear solution in terms of $\eta$, it satisfies the equations
\begin{align}
&\frac{d D}{d\eta} - f\mathcal{H}D = 0\,,\\
&\frac{d}{d\eta}(f\mathcal{H}D) + f\mathcal{H}^2 D - \frac{3}{2}\mathcal{H}^2\Omega_m D = 0\,.
\end{align}
The first of these is trivially satisfied using the definition $f \equiv d\ln D/d\ln a$. The second gives us a useful relation when changing variables from $\eta$ to $D$. In this case the linear equation can be written as
\be
\frac{d^2}{d D^2}\delta^{\iso} + \frac{3}{2}\frac{\Omega_m}{f^2} \frac{1}{D}\frac{d}{d D}\delta^{\iso} - \frac{3}{2}\frac{\Omega_m}{f^2}\frac{1}{D^2}\delta^{\iso} = 0\,.
\ee
If we now write $f=\Omega_m^{1/2}$, we obtain the growing mode $D_+(D) = D$ as should be, and the decaying mode $D_-(D) = D^{-3/2}$.

With these growing and decaying modes, we can write the function appearing in the evolution integrals as
\be
\mathcal{G}(a,s) = \frac{D_-(a) D_+(s) - D_+(a)D_-(s)}{D_+(s)\dot{D}_-(s) - D_-(s)\dot{D}_+(s)} = \frac{2}{5}\frac{s}{f}\left(D_+(a)D_+^{-1}(s) - D_+^{-3/2}(a)D_+^{3/2}(s)\right)\,,
\ee
which leads to the separable kernel (\ref{F2isoApprox}).

\section{Quadrupolar coefficients}\label{app:quadrupolar_coefs}
Here we give the quadrupolar tensors appearing in Eq.~(\ref{eq:F2sigma}). They can be obtained from a straightforward computation, and are given by:
\begin{align}
\begin{split}
A_{ij}(a) & = \int^a_{a_i} ds\,\mathcal{G}(a,s)\frac{\sigma_{ij}(s)}{s^2H(s)}\Bigg[\dot{D}_+(s)D_+(s)\\ &\phantom{deadwhitespace} - \int^s_{s_i} dh\, \frac{\partial}{\partial s}\mathcal{G}(s, h)\left(\frac{3\Omega_m^0}{2h^5}\left(\frac{H_0}{H(h)}\right)^2D_+(h)^2 + \dot{D}_+^2(h)\right)\Bigg]\,,
\end{split}  \\
B_{ij}(a) &= \int^a_{a_i} ds\,\mathcal{G}(a,s)\left[2\lambda_{ij}(s)\dot{D}^2_+-\frac{2\sigma_{ij}(s)}{s^2 H(s)}\int^s_{s_i} dh\, \frac{\partial}{\partial s}\mathcal{G}(s, h)\dot{D}_+^2(h)\right]\,, 
\end{align}

\begin{align}
\begin{split} 
C_{ij}(a) &= \int^a_{a_i} ds\,\mathcal{G}(a,s)\bigg[\frac{3\Omega^0_m}{2s^5}\left(\frac{H_0}{H(s)}\right)^2 D_+(s)(D_{+}(s)Q_{ij}(s)-D_{+}(s)\lambda_{ij}(s)) \\
&\phantom{deadwhitespace}+ \dot{D}_+(s)\left(-\dot{D}_+(s)\lambda_{ij}(s) + D_+(s)\dot{Q}_{ij}(s) + \dot{D}_+(s)Q_{ij}(s)\right)\\
&\phantom{deadwhispace}-\frac{2\sigma_{ij}(s)}{s^2H(s)}\dot{D}_+(s)D_+(s)\bigg]\,,
\end{split} \\
\begin{split}
E_{ij}(a) &= \int^a_{a_i} ds\,\mathcal{G}(a,s)\bigg[\frac{3\Omega^0_m}{2s^5}\left(\frac{H_0}{H(s)}\right)^2 D_+^2(s)\lambda_{ij}(s) \\ &\phantom{deadwhitespace}+\dot{D}_+(s)\left(\dot{D}_+(s)\lambda_{ij}(s) + \frac{1}{s^2H(s)}D_+(s)\sigma_{ij}(s) \right)\bigg]\,,
\end{split}\\
H_{ij}(a) & = \int^a_{a_i} ds\,\mathcal{G}(a,s)\left[-2\lambda_{ij}(s)\dot{D}^2_+(s) + \dot{D}_+(s)D_+(s)\dot{Q}_{ij}(s) 
+ \dot{D}^2_+(s)Q_{ij}(s)\right]\,, \\
K_{ij}(a) & = \int^a_{a_i} ds\,\mathcal{G}(a,s)\lambda_{ij}(s)\dot{D}^2_+(s)\,.
\end{align}

In the above expressions, a time-dependent $\lambda_{ij}$ has a physical effect. It can be written in terms of $\sigma_{ij}$ through 
\[
\lambda_{ij}(s)=\int\sigma_{ij}(s')ds'+\lambda^0_{ij}\,.
\]
With the exception of $A_{ij}(a)$, all tensors above contain terms linear in $\lambda_{ij}$. This means that if we naively take the isotropic limit as $\sigma_{ij}\rightarrow0$, terms proportional to $\lambda^0_{ij}$ remain, and so does $F^\sigma_2$, which might seem unexpected. However, note that all $\lambda^0_{ij}$ terms come from  the anisotropic corrections of the kernels $\alpha(\bfk_1,\bfk_2)$ and $\beta(\bfk_1,\bfk_2)$ in (\ref{alphabeta}). Indeed, starting from the definition of the kernels in a FLRW metric and promoting a constant rescaling of the coordinates, $x^i\rightarrow [\delta_{ij}+2\lambda^0_{ij}]x^j$, it is straightforward to check that the isotropic kernels get corrections given precisely by (\ref{deltaalpha}) and (\ref{deltabeta}). In other words, such terms are not observable, and disappear when working in a locally isotropic observational frame.

\bibliographystyle{JHEP} 
\bibliography{BiblioCPT} 
\end{document}